\documentclass[11pt]{article}
\pdfoutput=1

\usepackage{mathrsfs}

\usepackage{setspace}
\usepackage[letterpaper,top=25mm,left=31mm,bottom=28mm,right=31mm]{geometry}
\usepackage{titlesec}
\titleformat*{\section}{\Large\bfseries}
\titleformat*{\subsection}{\large\bfseries}
\titleformat*{\subsubsection}{\bfseries}
\titleformat*{\paragraph}{\bfseries}



\usepackage{soul}

\usepackage{mathtools}
 \usepackage{tcolorbox}
    \tcbuselibrary{listings,theorems}

\usepackage{upgreek}

\usepackage{empheq}
\usepackage{microtype}
\usepackage{relsize}
\usepackage[dvipsnames]{xcolor}
\usepackage{cite}
\usepackage{enumitem}
\usepackage{ifpdf}
\usepackage{lmodern}
\usepackage[T1]{fontenc}
\usepackage[ansinew]{inputenc}
\usepackage{amsmath}
\usepackage{amsthm}
\usepackage{relsize}
\usepackage{amssymb}
\usepackage{comment}
\usepackage{tikz}
\usepackage{graphicx}
\usepackage{color}
\definecolor{darkblue}{cmyk}{0.9,0.9,0,0}
\definecolor{darkgreen}{rgb}{0,0.55,0}
\usepackage[colorlinks=true,linkcolor=darkblue,citecolor=darkblue,urlcolor=darkblue]{hyperref}
\usepackage{epsfig}
\usepackage{epstopdf}
\usepackage{graphicx}

\makeatletter
\long\def\@makecaption#1#2{
  \vskip\abovecaptionskip
  \sbox\@tempboxa{{\captionfonts #1: #2}}
  \ifdim \wd\@tempboxa >\hsize
    {\captionfonts #1: #2\par}
  \else
    \hbox to\hsize{\hfil\box\@tempboxa\hfil}
  \fi
  \vskip\belowcaptionskip}
\makeatother

\def\j{{(\bJ)}}
\def\bJ{{\boldsymbol J}}
\def\cY{\mathcal{Y}}
\def\scL{\mathscr{L}}
\def\cL{\mathcal{L}}
\def\sfU{{\sf U}}
\def\pg{\paragraph}

\def\cI{\mathcal{I}}

\def\EE{\mathbb{E}}
\def\PP{\mathbb{P}}
\def\bitem{\begin{itemize}}
\def\eitem{\end{itemize}}
\def\benum{\begin{enumerate}}
\def\eenum{\end{enumerate}}

\definecolor{greencirc}{rgb}{0, 0.73, 0.5}

\def\enumcom{\begin{enumerate}[label*=  \sf (\roman*),wide, labelwidth=!, labelindent=0pt]}

\def\ni{\noindent}

\def\ZZ{\mathbb{Z}}

\def\RR{\mathbb{R}}

\def\vareps{\varepsilon}

\def\cF{\mathcal{F}}

\def\half{{1\o2}}

\def\x{\times}

\def\cD{\mathcal{D}}

\def\eps{\epsilon}

\def\hb{\overline h}
\def\Z{\mathbb{Z}}

\def\1{{\rm 1-loop}}

\def\Tr{{\rm Tr}}

\def\c{\cite}

\def\cA{\mathcal{A}}

\def\cM{\mathcal{M}}

\def\cN{\mathcal{N}}
\def\cA{\mathcal{A}}

\def\c{\cite}

\def\G{\Gamma}
\def\p{\partial}
\def\o{\over}
\def\g{\gamma}
\def\D{\Delta}
\def\rar{\rightarrow}

\def\eqr{\eqref}

\def\i{\infty}

\def\ssec{\subsection}

\def\sec{\section}

\def\foot{\footnote}
\newcommand{\es}[2] {\begin{equation} \label{#1} \begin{split} #2 \end{split} \end{equation}}
\newcommand{\e}[2] {\begin{equation} \label{#1} #2 \end{equation}}

\newcommand{\beq}{\begin{equation}}
\newcommand{\eeq}{\end{equation}}
\newcommand{\beqy} {\begin{eqnarray}}
\newcommand{\eeqy} {\end{eqnarray}}
\newcommand{\bsmat}{\begin{smallmatrix}}
\newcommand{\esmat}{\end{smallmatrix}}
\newcommand{\bmat}{\begin{matrix}}
\newcommand{\emat}{\end{matrix}}

\def\({\left(}
\def\){\right)}
\def\[{\left[}
\def\]{\right]}

\def\<{\langle}
\def\>{\rangle}

\def\a{\alpha}
\def\b{\beta}
\def\g{\gamma}
\def\G{\Gamma}
\def\d{\delta}
\def\D{\Delta}
\def\z{\zeta}

\renewcommand{\l}{\lambda}
\def\l{\lambda}

\def\t{\tau}
\def\s{\sigma}

\usepackage{varioref}
\usepackage{makeidx}
\makeindex

\usepackage[english]{babel}
\usepackage{caption}
\usepackage{subfig}

\numberwithin{equation}{section}

\usepackage{tabularx}
 
\begin{document}

\begin{spacing}{1.15}

\begin{titlepage}
\vspace*{2cm}
\begin{center}

{\Large \bfseries Black Holes and Random Variables}

\vspace*{1cm}

Eric Perlmutter

\vspace*{2mm}

\textit{\small Institut de Physique Th\'eorique, Universit\'e Paris-Saclay, CEA, CNRS\\Orme des Merisiers, 91191, Gif-sur-Yvette Cedex, France}

\vspace{2mm}

\vspace*{0.5cm}

\end{center}

\begin{abstract}

We formulate an avatar of the Fyodorov-Hiary-Keating conjecture for black hole microstate counts in quantum gravity. By holography, this implies sharp bounds on interval counts of high-dimension primary operators in conformal field theory. The extremal fluctuations of these counts are characterized by a random variable, with a prescribed tail distribution. At large $N$, these order-one erratic fluctuations set a quantitative limit on the resolution of the semiclassical AdS gravitational path integral. Gaussian random models for state counts arise naturally in this context; we express the phenomenon of erratic $N$-dependence in AdS/CFT as a decorrelation property of these models. Our broader point is to suggest that AdS black hole microstate spectra and their field theory duals should exhibit the extreme value statistics of random matrices, lying in the universality class of Gaussian log-correlated fields. 

\end{abstract}

\end{titlepage}
\end{spacing}

\pagenumbering{roman}
\begin{spacing}{0.7}

\setcounter{tocdepth}{3}
\tableofcontents
\end{spacing}

\pagenumbering{arabic}
\setcounter{page}{1}

\begin{spacing}{1.1}

\sec{Introduction}

In a pair of seminal papers \c{FHK,FK}, Fyodorov-Hiary-Keating and Fyodorov-Keating formulated a set of profound conjectures at the interface of random matrix theory and number theory. This set, to which we collectively refer as the FHK conjecture, pertains to the extreme value distributions of characteristic polynomials of random matrices and of the Riemann zeta function on the critical line, unified by the universal statistical properties of log-correlated fields. There has since been remarkable progress in these directions, including proofs of the central core of the conjecture. 

The goal of this paper is to bring the FHK conjecture to bear on high-energy spectra in quantum gravity and conformal field theory. 

The Riemann zeta function $\z(s)$ has long been an inspirational toy model for physicists seeking to understand the emergence of random matrix statistics from a deterministic quantum chaotic system. This dates back to work of Montgomery \c{Montgomery1973}, who argued that the pair correlation of the non-trivial zeros of $\z(\half+it)$ at $t\rar\i$, high on the critical line, takes the sine kernel form obeyed by Gaussian unitary random matrix eigenvalues -- a conjecture which was later supported by striking numerics by Odlyzko \c{Odlyzko1987}. At large $t$, there are of order $\log t$ zeros within a unit interval. 

Usually, it is the zeros which physicists regard as the spectral data in this model, in the spirit of Hilbert and P\'olya. But one can equally consider the $\sim \log t$ local extrema within the interval, and ask instead: what is the physical analog of {\it their} distribution? Indeed, there is an entire value field within the interval, and one may wonder what is the analog of its shape. We may visualize this impressionistically as some unspecified map between quantum chaotic spectra and Riemann zeta fluctuations:
\begin{figure}[h]
\centering
{\subfloat{\includegraphics[scale=.38]{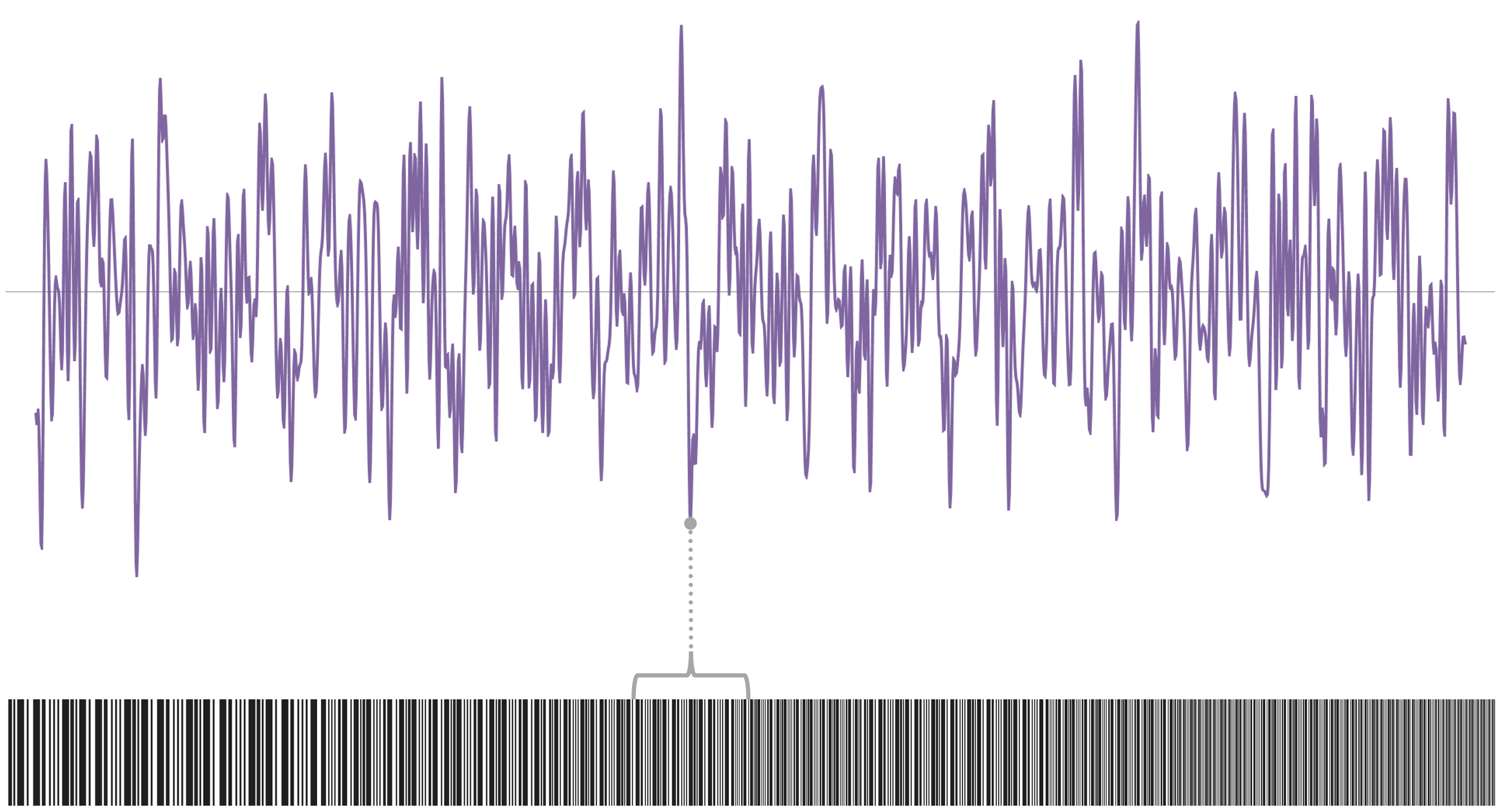}}}
\label{introfig}
\end{figure}

In quantum gravity, there are no black holes -- only black hole microstates. Accordingly, in quantum gravity, there is no black hole entropy -- only interval counts of these microstates. At the quantum level, for non-supersymmetric states, these counts remain difficult to study directly in the bulk. By the AdS/CFT correspondence, large black hole microstates map to high-dimension local operators in conformal field theory (CFT). Globally coarse-grained asymptotic counts of high-dimension CFT operators are well-studied using conformal bootstrap approaches \c{Cardy:1986ie, loga, shag, BLOSD}, but much less is established at the level of interval counts with rigorous control.  

While the original FHK conjecture was formulated with respect to the real parts of the spectral fields in question, as opposed to the discrete eigenvalue (or zero) counting staircase, some (well-known) massaging of imaginary counterparts of the conjecture \c{ABB15,PZ16,CMN,FLD} leads to versions that constrain the eigenvalue counts directly. As the counting staircase is native to AdS/CFT, it is this quantity -- the symmetry-resolved, mesoscopic interval counts of high-energy primary operators, dual to large AdS black hole microstates -- whose law we will consider. 

The title of this paper, a nod to \c{Cotler}, refers to a novel form of erratic behavior in AdS black hole spectra, more fine-grained than the spectral form factor: namely, the presence of a {\it random variable} that, following FHK, we postulate to characterize the {\it extreme spectral statistics} of large black hole microstates. The FHK conjecture is quite rich, with multiple layers, and their putative implications for black hole and CFT spectra are highly significant relative to what is known: as we will explain, even the leading order makes predictions for state counting that transcend what conformal bootstrap or other methods have established. 

More broadly, we aim to seed future explorations within AdS/CFT of the underlying ideas from extreme value theory of correlated systems and log-correlated statistics, which have received little attention in the high-energy theory community.

In Section \ref{sec2}, we review the FHK conjecture. We spend a bit of time developing the intuition for each layer of the conjecture, and state what has been proven in the literature. 

In Section \ref{sec3}, we review the ``imaginary FHK conjecture'' and state its implications for interval count fluctuations, which we denote as $\d \cN_I := \cN_I - \overline{\cN_I}$, where $\overline{\cN_I}$ is the mean density integrated over a spectral interval $I$. The conjecture may be viewed as a certain extension of a random Fourier model for $\d \cN_I$ with Gaussian-distributed coefficients which holds when the asymptotic spectrum obeys a central limit theorem with a (bulk) random matrix form factor. 

In Section \ref{sec4}, we state the AdS/CFT avatar of the FHK conjecture for interval counts. 

In Section \ref{sec5}, we present five points of interpretation, focusing mostly on the main statement for state counts in quantum gravity and in CFT. In quantum gravity, we emphasize the universality and origin of log-correlated statistics, deferring discussion of the spacetime interpretation to Section \ref{sec6} in which we discuss large $N$ and semiclassical AdS/CFT. In CFT, we recognize the leading-order FHK constraint on $\d\cN_I$ (proven in \c{HP,CMN,CFLW,BLZ}) as a bootstrap-type result for $d$-dimensional CFTs: it is an optimal bound for interval counts. This is a rigorous statistical counterpart of bootstrap approaches using Tauberian theorems \c{espin,RychkovYvernay2016,QiaoRychkov2018,MZ19,baursasha,MP,VanRees2025}. In fact, purely spectral Tauberian bounds are only known in $d=2$, so we view our result as significant. We explain how our result strengthens the existing $d=2$ bounds and addresses the goal of incorporating spectral statistics into the bootstrap. 

In Section \ref{sec6}, we analyze the conjecture in the large $N$ limit, in which the bulk is semiclassical AdS gravity. The main implication for semiclassical gravity is that, uniformly over mesoscopic intervals at high energy, extremal fluctuations around the mean are linear in the gravitational entropy $S_0 \approx \log\rho_0$, with a subleading irreducible stochastic fluctuation $\cY_\i$ of $O(1)$ amplitude. The random variable $\cY_\i$ is random in the position of the interval; it captures non-self-averaging contributions in energy. The average term $\overline{\cN_I}$ is standardly interpreted as the smooth contribution from the semiclassical gravitational path integral, of $O(e^{S_0})$. Consequently, the scale of the erratic fluctuations implies a quantitative precision limit for the Euclidean semiclassical gravitational path integral, of $O(e^{-S_0})$. This advances recent discussions on the limits of semiclassical gravity, by making a precise proposal for the latter in the microcanonical ensemble. We further note a relation to questions of off-shell contributions in semiclassical gravity, and discuss the relation to FZZT brane correlators in JT gravity. 

In Section \ref{sec61}, we back off from the FHK setting and make some remarks on the erratic $N$-dependence of holographic CFT data, in particular for the Gaussian random models of microstate counts $\d \cN_I$. The result is a CFT embedding of a mechanism for erraticness identified in \c{KFW} on the basis of AdS wormhole amplitudes. 

In two appendices, we explain how Gaussian random models arise from diagonalizing central limit theorems, and make some comments on the FHK conjecture in the broader context of universality in quantum chaos. 

\sec{FHK Conjecture}\label{sec2}

The goal of this section is to introduce the FHK conjecture, to briefly explain its meaning and significance, and to state what has been proven so far.  

\ssec{General shape}

All of the conjectures in this section and the next are of the following type. Consider a centered spectral field $X_{N,{\sf U}}(z)$, on a domain $\mathcal{D}$ containing $O(N)$ microscopic degrees of freedom in a statistical universality class ${\sf U}$. Then as $N\rar\i$, 
\e{fhkgen}{\sup_{z\in \mathcal{D}}X_{N,{\sf U}}(z)= C_{\sf U}\(\log N -{3\o 4}\log\log N + Y_{N,{\sf U}}\)}
where $Y_{N,\sfU}$ is a sequence of $O_\PP(1)$ random variables, converging in law to a limiting variable $Y_{\i,\sfU}$ with tail probability 
\e{piy}{\PP(Y_{\i,\sfU} > y) \sim ye^{-2y}}
as $y\rar\i$. The coefficient $C_{\sf U}$ encodes the universality class, and is independent of $N$. 

The source of randomness of $Y_{N,\sfU}$ depends on the specific setting: for deterministic spectra, probability is induced by uniformly sampling the basepoint of $\cD$ within a larger domain over which the local spectral density is stationary, whereas for random matrices, $\cD$ may be held fixed and the randomness comes from the ensemble draw. The domain $\cD$ is often taken to be either a macroscopic interval $\cI_\cL$ or a mesoscopic interval $I_L\subset \cI_\cL$.\foot{{\it Stationary} means that the mean density stays nearly flat across the interval. {\it Macroscopic}, or {\it global}, means that the interval contains an $O(1)$ fraction of all states within a given stationary block. {\it Mesoscopic} means that the window contains an $o(1)$ fraction of states within the block, but still an infinite number of states.} It is helpful to write this in unfolded coordinates $u=zN$. Taking $\cD = \cI_\cL$, for example,
\e{fhkgenunfold}{\sup_{u\in I_\scL}X_{N,\sfU}(u) = C_{\sf U}\(\log \scL -{3\o 4}\log\log \scL + Y_{\scL,\sfU}\)}
where $\scL=\cL N\rar\i$ is the unfolded length of the interval (and $Y_{\scL,\sfU}$ picks up a simple $\cL$-dependent shift). 

\ssec{Statement of conjectures}\label{sec22}

The original works of FHK presented two such conjectures.\foot{Actually, \c{FHK,FK} present further accoutrements of these conjectures, especially regarding the moments of characteristic polynomials and their statistical mechanical interpretation, that we will not present here.}

The first pertains to the characteristic polynomials of $N\x N$ Haar-distributed random matrices in the circular unitary ensemble (CUE). For matrices $M_N$ with $N$ eigenvalues $\{\l_j\}$, the characteristic polynomial is defined as
\e{}{P_N(z) := \det(z-M_N) = \prod_{j=1}^N(z-\l_j)}
In circular ensembles, $|\l_i|=1$. The original FHK conjecture for random matrices takes the field $X_{N,\sfU}(z)$ in \eqr{fhkgen} to be the log determinant for $\sfU =$ CUE, and can be stated as follows: 
\e{fhk1}{\sup_{|z|=1}\log |P_N(z)| = \log N -{3\o4}\log\log N + Y_N}
as $N\rar\i$, where $Y_N$ is a sequence of $O_\PP(1)$ random variables converging in law to a limiting random variable $Y_\i$ obeying \eqr{piy}. In fact, a precise conjecture was given for the limiting probability density \c{FB,FHK,FK}:
\e{pbessel}{p(y) = 4e^{-2y}K_0(2e^{-y})}
from which one easily ascertains the asymptotic scaling \eqr{piy}, with coefficient $2$.\foot{We note the relation $\p_y(2e^{-y} K_1(2e^{-y})) = 4e^{-2y}K_0(2e^{-y})$ and the $y\rar\i$ asymptotic $2e^{-y}K_1(2e^{-y}) \approx 1-2ye^{-2y}$. In the literature one often finds the probability distribution stated in terms of $-2\log|P_N(z)|$, which is a normalization convenient for the statistical mechanical interpretation of the behavior \eqr{fhk1} as a freezing transition. The probability densities are related as $p(y)|_{\rm here} =2p(-2y)|_{\rm there}$.} This admits slight generalization to a version where one extremizes over arcs $\theta:= \arg(z) \in [0,L]$ for $0<L<2\pi$; for any finite $L$, this is still a macroscopic interval because it contains an $O(1)$ fraction of all $N$ eigenvalues.

This was later directly extended to the circular $\b$-ensembles (C$\b$E), Gaussian ensembles, and Wigner ensembles (among others), and it will be useful to quote that here. We state it in the form presented in \c{BLZ}. Let $\cI$ be a fixed macroscopic interval supported in the bulk of the spectrum in the non-circular case, i.e. $\cI := \{z\in [-2+\eps,2-\eps]\}$ for a fixed $\eps>0$, or on a finite arc in the circular case, i.e. $\cI := \{\theta\in[0,L)\}$ for fixed $0<L<2\pi$. Then the conjecture is that 
\e{fhk2global}{\sup_{z \in \cI}\(\log |P_N(z)| - \EE[\log|P_N(z)|]\)= \sqrt{2\o\b}\(\log N -{3\o4}\log\log N + Y_{N,\b}\)}
as $N\rar\i$, where $Y_{N,\b}$ are sequences of $O_\PP(1)$ random variables converging in law to limiting random variables $Y_{\i,\b}$ obeying \eqr{piy}. The cases $\b=1,2,4$ correspond to orthogonal, unitary and symplectic cases, respectively. Importantly, for macroscopic intervals $\cI$, the respective distributions of $Y_{\i,\b}$ depend not only on $\b$ but also on the global eigenvalue structure, and hence can be expected to differ for the circular and non-circular cases: indeed, in the Gaussian ensembles, little is rigorously proven about $Y_{\i,\b}$ in the macroscopic case. 

The second conjecture of FHK pertains not to random matrices, but to the logarithm of the Riemann zeta function over short intervals on the critical line. The conjecture may be stated as follows. Define the short interval $I := [\t,\t+2\pi]$ and the stationary interval $\cI := [T,2T]$. For $\t \sim \text{Unif}(\cI)$, 
\e{fhkzeta}{\sup_{t \in I}\log\left|\z\(\half+it\)\right| = \log\log T - {3\o 4}\log\log\log T + Y_T}
as $T\rar\i$, where $Y_T$ is a sequence of $O_\PP(1)$ random variables converging in law to a limiting random variable $Y_\i$ whose law enjoys the asymptotic scaling \eqr{piy}. 

These conjectures admit extensions to mesoscopic intervals, of unfolded length of $O(N^\theta)$ in the random matrix case and of $O((\log T)^\theta)$ in the Riemann zeta case, with $\theta\in(0,1)$. The extremal laws take the same form, up to additive shifts of the $O_\PP(1)$ random variables by a Gaussian random variable with variance of $O(\log N)$ in the case \eqr{fhk2global} and of $O(\log \log T)$ in the case \eqr{fhkzeta} \c{FK}. These terms are ``coarse'' fluctuations common to every point in the interval, and indeed may be removed by subtracting the averaged field over the interval: in the notation of \eqr{fhkgen}, they would disappear if one considered the laws for the centered fields 
\e{29meso}{X_{N,\sfU}(z) - {1\o |\cD|} \int_\cD dz\, X_{N,\sfU}(z)}
Note that the Gaussian and circular mesoscopic laws have the same form, up to subleading $1/N$ corrections to the limiting law, provided that we locally unfold the Gaussian spectrum over a bulk stationary interval. 

\ssec{Meaning}\label{s23}

Let us explain how to think about this general class of conjectures \eqr{fhkgen} -- why they ought to be true, and what physics they contain -- after which we make some remarks specific to the Riemann zeta conjecture \eqr{fhkzeta}. This is meant to build intuition, not as a review; the reader may find a comprehensive, relatively recent overview in \c{BK}. 

There are three layers to these conjectures, each signaling a different conceptual paradigm:
\begin{figure}[h]
\centering
{\subfloat{\includegraphics[scale=.52]{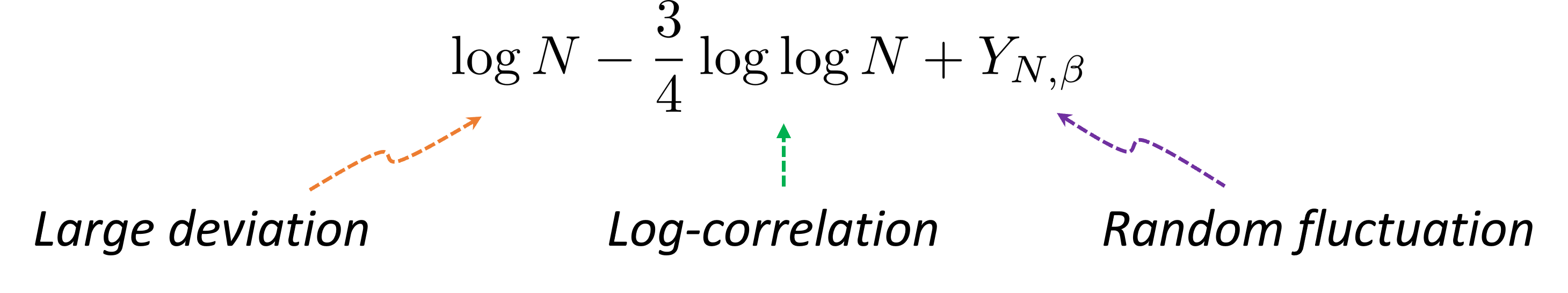}}}
\label{123pic}
\end{figure}

\noindent \textbf{Leading order (large deviation):} The leading-order term may be viewed through the lens of {\it large deviation theory}. A typical starting point is to assume that the spectral point process in question obeys a mesoscopic central limit theorem (CLT), more precisely, that the observable $X_{N,\sf U}(z)$ obeys a mesoscopic CLT. This means that upon appropriate statistical sampling over mesoscopic intervals,
\e{}{X_{N,\sfU}(z) \Longrightarrow \cN(0,V_N)}
with variance $V_N := V_N(z)$ (we henceforth suppress $z$- and $\sfU$-dependence of the variance, and $\sfU$-dependence of $X_{N,\sfU}(z)$). For the log-characteristic polynomials $\log |P_N(z)|$ of random matrices, this holds with variance \c{KSzeta}
\e{}{V_N \approx {1\o\b}\log N}
for fixed $z$. This means that the typical value of $X_{N}(z)$ is $O(\sqrt{\log N})$. However, the question of the supremum of $X_{N}(z)$ at scales far greater than the standard deviation is not governed by the CLT. The domain of such rare excursions is that of {\it large deviation principles} and {\it extreme value theory}. 

Nevertheless, the CLT provides a useful heuristic with which one can derive the leading $\log N$ term by extrapolation of the CLT to large scales, as follows.\foot{A random variable obeying suppression of large deviations at scales $y \gg \sqrt{V_N}$ is said to satisfy a large deviation principle (LDP). When the falloff is Gaussian, this is known as Gaussian concentration.} Assume that within the mesoscopic interval, the $N$ levels are independent. We can then estimate the supremum $y_*$ as the solution to
\e{ld1}{N\,\PP(X_{N}(z) > y_*) = 1}
If we naively extrapolate the real Gaussian falloff $\PP(X_{N}(z)> y) \sim e^{-y^2/2V_N}$ out to large scales $y \gg \sqrt{\log N}$, we find
\e{ld2}{y_* \approx \sqrt{2V_N\log N}}
Since $V_N \approx \b^{-1} \log N$ in RMT, the supremum $y_*$ is simply linear in the variance. This reproduces the leading term $\sqrt{2\o \b}\log N$ in \eqr{fhk2global}, including the coefficient. 

Now, if one keeps the fluctuation determinant factor in the above computation by taking $\PP(X_{N}(z) > y) \sim e^{-y^2/2V_N}/\sqrt{V_N}$, one obtains a subleading correction
\e{}{y_* \approx \sqrt{2\o\b}\(\b V_N-{1\o 4}\log V_N\)}
as $V_N\gg1$. Comparing to \eqr{fhk2global}, this gives a coefficient of $-{1\o4}$ instead of $-{3\o4}$. The reason is that in the cases of interest to FHK, our assumption of independence of the $N$ levels fails: the field $X_{N}(z)$ is, in fact, {\it logarithmically-correlated}. 
\vskip .1in
\noindent\textbf{Subleading order (log-correlation\foot{(a.k.a. another famous ${3\o4}$ \c{peet})}):} An intuitive way to understand the log-correlated structure is by upgrading the Gaussian concentration in probability to a field realization, as a random Fourier series representation of $X_N(z)$. For cases of interest, in which the variance of $X_N(z)$ is logarithmic, this takes the general form
\e{}{X_N(z) \propto\Re\[\sum_{k\leq N} g_k{e^{2\pi i k z}\o \sqrt{k}}\]}
where the $g_k$ are complex Gaussian random variables with $\EE[g_k\overline g_{k'}] = \d_{kk'}$. This can be understood as formally diagonalizing the Gaussian fluctuations of a CLT with logarithmic variance, and promoting them to a field; we explain this in Appendix \ref{appA}. For the example of the centered log-characteristic polynomial of a $\b$-ensemble, it admits a representation 
\e{rfmp}{X_N(z) := \log|P_N(z)| - \EE[\log|P_N(z)|] = \sqrt{2\o\b} \Re\[\sum_{k\leq N} g_k{e^{2\pi i k z}\o \sqrt{k}}\]}
where we have discretized momenta over a unit interval, i.e. of unfolded length $N$. (This representation is actually exact in the CUE at large $N$, where $g_k = \lim_{N\rar\i}{1\o \sqrt{k}} \Tr(U_N^k)$ are asymptotically Gaussian-distributed.) Computing the covariance from \eqr{rfmp},
\e{}{\EE[X_N(z)X_N(z')] = {1\o\b} \sum_{1\leq k \leq N} {\cos(2\pi k (z-z'))\o k}}
For $|z-z'|$ fixed and $N\rar\i$, 
\e{EEXX}{\EE[X_N(z)X_N(z')] = -{1\o\b}\log\(2\,|\!\sin(\pi(z-z'))|\)}
Backing off slightly by treating $N$ as a cutoff, one finds the following behavior of the sum:
\e{EEXX2}{\EE[X_N(z)X_N(z')]  \approx {1\o\b}\x\begin{cases} 
-\log|z-z'|\qquad {1\o N} \ll |z-z'| \ll 1\\ \log N \qquad\qquad |z-z'| \ll {1\o N}\end{cases}}
One observes logarithmic correlation above all scales $|z-z'| \sim 1/N$: beyond the mean level spacing, the covariance decreases by one unit for every $e$-fold of separation $|z-z'|$. 
This hierarchical structure can be seen more explicitly by grouping the Fourier modes into dyadic shells or $e$-folds,
\e{Xr}{X_r(z) := \sqrt{2\o\b}\sum_{e^{r-1}\leq  k < e^r}\Re\[g_k{e^{2\pi i k z}\o \sqrt{k}}\]}
such that
\e{}{X_N(z) =\sum_{r=1}^{\log N} X_r(z)}
The variance of one $e$-fold $X_r(z)$ is $O(1)$, and there are approximately $\log N$ shells. The shells with $r \gg -\log |z-z'|$ are effectively decorrelated. This multiscale behavior, in which nearby points share scales $r \lesssim -\log |z-z'|$, is that of a branching random walk \c{Arguin2017}. 

These nested correlations can be shown to lead to a $-{3\o4}\log V_N$ correction instead of a $-{1\o4}\log V_N$. In our heuristic large deviation computation \eqr{ld1}, the mechanism that modifies the probability is the imposition of a so-called ``barrier condition,'' which amounts to an extra suppression by a barrier probability $V_N^{-1}$. So one instead solves
\e{}{N \,\PP(X_N(z) > y_*) V_N^{-1} = 1}
from which one easily extracts the $-{3\o4} \log V_N$ \c{Bramson1978}. The barrier imposes the fact that an extremal path must remain viable across all dyadic scales $X_r(z)$, not merely end high at the final scale. In other words, the fact that the supremum should be lower is intuitively clear: the correlations make it ``harder'' to achieve higher values, because the field must follow a whole extremal path in $z$ by climbing a landscape of local suprema.

We stress that the random Fourier model is a field ansatz for the Gaussian central fluctuations, log-correlated due to spectral rigidity, that is then being extrapolated to describe large deviations. This extrapolation accurately captures the extreme statistics to subleading order, for random matrices and their universal limiting point processes, or any deterministic process\foot{For deterministic processes, Fourier coefficients of the unfolded density fluctuation in a window are just fixed numbers. The random Fourier model replaces those coefficients by Gaussian variables because this is what it means to obey a CLT: their statistics as a function of the location of the window are approximately Gaussian.} whose statistics converge appropriately thereto.

\pg{Random fluctuation:} The way to really read \eqr{fhkgen} is as a statement of convergence in law after appropriate centering: as $N\rar\i$, 
\e{}{C_{\sf U}^{-1}\sup_{z\in \mathcal{D}}X_{N,{\sf U}}(z)- \(\log N -{3\o 4}\log\log N\) \Longrightarrow Y_{\i,{\sf U}}}
The identification of limiting laws for extreme values of random processes is precisely the domain of {\it extreme value theory}: do the extrema, after a deterministic subtraction, themselves converge in law? If so, to what? 

The presence of correlations, as in the log-correlated universality class being considered here,  necessitates a departure from classical extreme value theory -- in which limiting laws fall into one of three canonical classes \c{DemboZeitouni1998} -- to a more general domain of extreme value theory for correlated systems.\foot{See \c{Arguin2017} for an accessible introduction to the subject.} A hallmark of log-correlated statistics in particular is the tail asymptotic \eqr{piy}. Compared to a (rescaled) Gumbel distribution, which has probability density function $p(y) = \exp(-y -e^{-y})$, \eqr{piy} is dressed by an extra power of $y$, immediately signaling that $Y_{\i,\sfU}$ cannot simply be Gumbel. Indeed, the conjectural Bessel form \eqr{pbessel} was later identified as the sum of two i.i.d. Gumbels \c{FHK,sabag}.

The random process contains more information beyond the limiting law of its suprema. $Y_{\i,\sfU}$ describes just a top slice through an entire {\it extremal landscape}: a limiting marked extremal point process, comprised of the heights and locations of the local maxima decorated by the shape of the distribution in the local neighborhood of these points. While the shape of this landscape goes beyond the original FHK conjectures per se \c{PZ22},  those conjectures were centrally motivated by statistical mechanical considerations of disordered landscapes, and their so-called freezing transition. Defining an auxiliary ``extremal'' partition function and free energy as
\e{freeenergy}{Z_{\rm ext}(\b)  := \int_{\cD} d\mu(z)e^{\b X_N(z) }\,,\quad \cF_{\rm ext}(\b) := {1\o \b} \log Z_{\rm ext}(\b)}
where $\mu(z)$ is the measure over $\cD$, then 
\e{}{\sup_{z \in \cD} X_N(z) = \lim_{\b\rar\i} \cF_{\rm ext}(\b)}
The freezing transition is a phase transition at $\b=\b_c$ above which $\cF_{\rm ext}(\b)$ flattens: the measure effectively condenses onto the atoms of the extremal point process. This glassy perspective is explained at length in the original papers \c{FHK,FK} to which we refer the reader for more details.

\pg{Riemann Zeta conjecture:}  

The FHK conjecture \eqr{fhkzeta} for $\log|\z(\half+it)|$ is a valuable example of an extremal law for a {\it deterministic} spectrum, induced by statistical sampling.

The FHK conjecture \eqr{fhkzeta} may be read as the assertion that $\log |\z(\half+it)|$ has the extreme value statistics of a GUE log-characteristic polynomial, and therefore, in view of \eqr{fhk1}, that its extremes also lie in the universality class of log-correlated fields. The conjecture \eqr{fhkzeta} was a non-trivial extension of previously-established GUE-type properties of Riemann zeta to the large deviation regime: indeed, data from Selberg's CLT for $\log\z(\half+it)$ with variance $V \approx \half \log\log T$, the pair correlation \c{Montgomery1973, Odlyzko1987}, the $2k$'th critical moments and the value distribution \c{KSzeta} indicated that over short intervals high on the critical line, Riemann zeta behaves essentially like a GUE random matrix with the identification $N := \log{T\o 2\pi}$ as $T\rar\i$.\foot{We make two technical remarks. First, the choice of the specific interval $[T,2T]$ in \eqr{fhkzeta} is not special. The intended statement is just that a uniformly sampled point high on the critical line sees a typical local log-correlated environment, and $[T,2T]$ is just a canonical way of implementing that sampling. Second, the $T\rar\i$ limit is taken first, so \eqr{piy} is not valid for $y$ growing arbitrarily fast in $T$ (and likewise for \eqr{fhk1} in $N$). The question of how far the scaling \eqr{piy} may be extrapolated connects to the question of the global maximum of $|\z(\half+it)|$ \c{gonek,Sound22}.} 

A complementary view of the FHK conjecture for $\log|\z(\half+it)|$ is as an instantiation of a general expectation, nicely articulated in \c{ABB15}, that log-correlated fields obey universality even at the level of extreme value statistics, in particular, with the prescribed centering term and the limiting law tail asymptotic \eqr{piy}. It was first observed in \c{Bourgade2009} that $\log|\z(\half+it)|$ exhibits logarithmic correlation and branching phenomena, and then again in \c{FHK,FK}, providing independent evidence for \eqr{fhkzeta} of this type. Numerical evidence for \eqr{fhkzeta} through subleading order was also provided in \c{FHK,FK}.

\ssec{State of the art}\label{sec24}
We give a compact account of the status of the conjectures \eqr{fhk1}, \eqr{fhk2global} and \eqr{fhkzeta} (see \c{BK} for a more complete history). Essentially, the circular ensemble and Riemann zeta conjectures have been proven modulo the explicit form and finer properties of the limiting distributions, while the non-circular ensemble conjectures have only been proven to leading order.

Beginning with the CUE random matrix conjecture \eqr{fhk1}, the leading-order term was proven by \c{ABB15} which also established the associated freezing transition for the free energy \eqr{freeenergy}. The second-order term was proven by \c{PZ16}. Soon thereafter, the work of \c{CMN} established stronger results, proving tightness of the centered maximum for the C$\b$E, and for both real and imaginary parts (to be discussed immediately below). The C$\b$E story was advanced further in \c{PZ22}, which characterized the extremal landscape and proved convergence in law of the centered maxima to a randomly shifted Gumbel variable. The precise law for $Y_{\i,\b}$ remains a subject of active research. 

Concerning the extension of the macroscopic FHK law \eqr{fhk2global} to non-circular ensembles, \cite{FyodorovSimm2016} formulated the corresponding conjecture for the maximum of the bulk GUE characteristic polynomial, the leading-order of which was proven by \cite{Lambert2019}, also for general one-cut unitary-invariant Hermitian ensembles (see also \cite{LambertPaquette2023Hyperbolic,LambertPaquette2024Edge} for Gaussian $\b$-ensembles). Proof of leading-order scaling was done for the more general Wigner matrices and general $\b$-ensembles by \c{BLZ}, unifying and extending the previous results, and again for both the real and imaginary parts. Relatedly, the maximum deviation of the Sine$_\b$ counting process and optimal eigenvalue rigidity have been established at leading order in \cite{HP,CFLW}.\foot{While we have not discussed it here, \c{FHK,FK} also make predictions for moments of characteristic polynomials. See e.g. \c{BaileyKeating2019Moments,AssiotisKeating2020,BK} for further details and references. We also note in passing that the modern viewpoint on the FHK conjecture, and in particular on the rigorous probabilistic formulation of its statistical mechanical interpretation, has evolved to include techniques from Gaussian multiplicative chaos, e.g. \c{Webb2015, NikulaSaksmanWebb2020,LambertNajnudel2026}.}

Passing now to the Riemann zeta conjecture \eqr{fhkzeta}, various works proving parts of the conjecture to leading \c{ABBRS,Najnudel} and subleading \c{Harper2019Partition} order, as well as the study of analogous behaviors of random models of Riemann zeta \c{Harper2013, ArguinBeliusHarper2017}, culminated in the works \c{ABR1,ABR2} which proved tightness of the centered maximum of $\log|\z(\half+it)|$ over short intervals, and moreover established the tail asymptotic \eqr{piy} over large ranges of $y$ scaling with $t$. As in the random matrix case, the precise limiting law is unknown. Finally, the mesoscopic case has been proven to a slightly lesser degree in \c{ArguinOuimetRadziwill2021, ArguinHamdan2024}.

\sec{Imaginary FHK and Counting Statistics}\label{sec3}

Noting that $\log|P_N(z)| = \Re\log P_N(z)$, and that $\log P_N(z)$ is a complex field, it is natural to formulate extreme value conjectures of the form \eqr{fhk2global} for the imaginary parts $\Im\log P_N(z)$ in random matrix theory, and of the form \eqr{fhkgen} for imaginary parts of complex fields $X_{N,\sfU}(z)$ more generally. Though by now recognized\foot{While not strictly part of the original FHK conjectures, the imaginary extension was, as far as we are aware, first proposed as a natural analog in \c{ABB15} to leading-order, and in \c{PZ16,CMN} to subleading order, specifically for unitary matrices. Later work developed this further \c{HP, CFLW, BLZ, FLD,PZ22}.} as part of the suite of log-correlated extreme value conjectures for random matrices, ``imaginary FHK'' is most often regarded as a secondary corollary, or as a stepping stone to results on optimal rigidity of eigenvalue spacings. However, for our purposes, these are paramount: extremal value conjectures for the imaginary parts of log-determinants may be recast in terms of {\it interval counts of eigenvalues}. These are fundamental observables in the setting of AdS quantum gravity and non-perturbative conformal field theory.

Let us begin in the setting of log-determinant operators. Define the centered imaginary field
\e{}{\Phi(z) := \Im\log P_N(z) - \EE[\Im\log P_N(z)]}
The analogous FHK conjecture for $\Phi(z)$ is again of the functional form \eqr{fhk2global}, but with an important difference: the bounds in the literature \c{HP,CMN,CFLW,BLZ} are for both the {\it supremum and infimum of $\Phi(z)$}. We write them in a unified notation as
\e{imfhk}{\qquad \sup_{z \in \cD}\(\s\Phi(z)\)= \sqrt{2\o\b}\(\log N -{3\o4}\log\log N + Y_{N,\b}^{(\s)}\)\qquad (\s = \pm)}  
as $N\rar\i$. (The domain $\cD$ may be either a mesoscopic or macroscopic interval with respective random variables $Y_{N,\b}^{(\s)}$, keeping in mind the distinction noted in Section \ref{sec22}.) The sequences of random variables $Y^{(+)}_{N,\b}$ and $Y^{(-)}_{N,\b}$ characterize the suprema and infima, respectively, whose limiting laws again have the log-correlated tail \eqr{piy}, though no conjectural analog of \eqr{pbessel}.\foot{In the notation of \c{PZ22}, $Y_{\i,\b}^{(\pm)} = \sqrt{\b\o8}(a_\i \pm I_\pm)$ where $a_\i$ is a subsequential limit of a sequence of deterministic constants $a_N \in [0,2\pi]$.} The leading-order statement has been proven for both mesoscopic and global intervals, in myriad ensembles (circular, Wigner, $\b$) and for Sine$_\b$  \c{HP,CMN,CFLW,BLZ}. At the subleading order, there are results only for the circular ensembles: for C$\b$E, tightness to subleading order was proven in \c{CMN} for all $N \geq 2$, while convergence in law as $N\rar\i$ was later proven in \c{PZ22}. 

Now, following these works, let us pass from the field $\Phi(z)$ to interval counts. Following the standard logarithmic branch convention therein, the imaginary part of the log is an indicator function, $\Im\log(z-\l) = \pi\mathbf{1}_{\{\l>z\}}$. 
By definition of the log-determinant, this implies
\e{}{\Phi(z) = \pi \Bigg(\sum_j \mathbf{1}_{\{\l_j>z\}}-\int_z^\i dx \,\rho(x)\Bigg)}
where $\rho(x)$ is the coarse-grained density of states for the system in question. Taking a difference, 
\e{counts}{\Phi(z) - \Phi(z') = \pi\(\cN_{[z,z']} - \overline \cN_{[z,z']}\)~~ \text{where}~~\overline \cN_{[z,z']} := \int_z^{z'} dx\, \rho(x)}
Thus, up to an overall factor of $\pi$, the difference $\Phi(z) - \Phi(z')$ is simply the number fluctuation 
\e{}{\d \cN_{[z,z']} := \cN_{[z,z']} - \overline \cN_{[z,z']}}
around the average count $\overline\cN_{[z,z']}$ in the interval $[z,z']$. In unfolded coordinates, we write $\d\cN_{[u,u']}$. Note that for a stationary interval, $\overline \cN_{[u,u']}$ is simply the length of the interval in unfolded coordinates. 

Straightforwardly combining the above ingredients yields a version of the FHK conjecture for interval counts of random matrix eigenvalues. Before considering the supremum and infimum of $\d \cN_{[z,z']}$ separately, it is simpler to consider the extremal law for the absolute value $|\d \cN_{[z,z']}|$. In view of \eqr{counts}, its supremum is $\pi^{-1}$ times the {\it range} of the field $\Phi(z)$ for $z\in\cD$,
\e{doublesup}{\sup_{z,z' \in \cD}\(\Phi(z)-\Phi(z')\) = \sup_{z\in \cD}\Phi(z) -  \inf_{z\in \cD}\Phi(z)}
By \eqr{imfhk}, the centering terms simply double: the corresponding extremal law is thus
\e{Range'}{\sup_{[z,z']\in \cD} |\d\cN_{[z,z']}| = {1\o \pi}\sqrt{2\o\b}\[2\(\log N -{3\o4}\log\log N\) + \cY_{N,\b}\]}
as $N\rar\i$, with $O_\PP(1)$ random variables $\cY_{N,\b}$. 

The natural next question is that of the tail asymptotic of $\cY_{\i,\b}$, analogous to \eqr{piy}. The answer, which follows from \eqr{doublesup}, is that the limiting law is determined by the {\it joint} law of infima and suprema of $\Phi(z)$. Let us investigate this a bit more explicitly. Parameterizing the tail scaling as
\e{piy2}{\PP(\cY_{\i,\b} > y) \sim y^pe^{-2y}}
as $y\rar\i$ for some $p\in\RR$, the exponential is the dominant feature of log-correlated statistics; as we now explain, a natural expectation for the power law is $p=3$. According to Conjecture 1.13 of \c{PZ22}, joint convergence\foot{What can fail in principle is the existence of a limiting law: different couplings between the positive and negative extremes can give different subsequential laws. The content of Conjecture 1.13 of \c{PZ22} is that this does not happen. In any case, tightness of $\cY_{N,\b}$ is guaranteed by tightness of $\sup\limits_{z\in \cD}\Phi(z)$ and $\inf\limits_{z\in \cD}\Phi(z)$ after subtraction \c{PZ22}.} holds, which would imply that $\cY_{N,\b}$ converges in law to the sum
\e{}{\cY_{N,\b} \Longrightarrow Y_{\i,\b}^{(+)}+Y_{\i,\b}^{ (-)}}
If the joint tails of $Y_{\i,\b}^{(\pm)}$ are independent, then the tail of $\cY_{\i,\b}$ is determined by their convolution, which gives $p=3$:
\e{}{\int_{y_0}^y dx (x e^{-2x})\big((y-x)e^{-2(y-x)}\big) \approx {1\o 6}y^3e^{-2y}}
as $y\rar\i$. If the tails are not independent, the question is whether their correlation is sufficiently strong to generate $p\neq 3$. In the same spirit as the statistical mechanical motivations of \c{FHK,FK} for the original FHK conjectures, one would also like a separate argument as to the expected value of $p$ from log-correlated universality, not using assumptions about joint convergence per se. Precisely this is provided by \c{cld}, who studied the range statistic in the so-called ``circular $1/f$-noise'' model, a Gaussian log-correlated field: indeed, they found that the min-max correlation {\it is} nonzero, but only changes the overall coefficient of the tail, preserving $p=3$ (cf. eq. (33) and footnote 5 of \c{cld}).

So, summarizing the above, the asymptotic tail probability \eqr{piy2} of $\cY_{\i,\b}$ in the FHK law for $|\d\cN_{[z,z']}|$ is determined by the joint law of maxima and minima, with $p=3$ a natural baseline expectation. We emphasize that a direct probabilistic derivation remains open (as for the tail \eqr{piy}).

Extremal laws of the same form as \eqr{Range'} actually hold for the supremum and infimum of $\d\cN_{[z,z']}$ separately. On the one hand, the supremum over ordered intervals $[z,z'] \in \cD$ with $z<z'$ is, in general, only upper-bounded by ($\pi^{-1}$ times) the full range of $\Phi(z)$: in particular, the maximum of $\Phi(z)$ over $\cD$ could occur to the {\it right} of the minimum. In other words, the supremum is fixed by the maximum ``drawdown'' of $\Phi(z)$ over $\cD$. On the other hand, a simple adornment of the argument leading to \eqr{Range'} yields an extremal law of the same form for the supremum: splitting $\cD$ into two fixed subintervals each of which occupies an $O(1)$ fraction of $\cD$, and applying the FHK law on each, it follows that the centered supremum is tight with respect to the same centering as in \eqr{Range'}. The exact same argument applies to the infimum of $\d\cN_{[z,z']}$, which is fixed by the maximum ``drawup'' of $\Phi(z)$ over $\cD$. 

Therefore, the signed FHK law for interval counts is
\e{Range}{\qquad \sup_{\substack{z,z' \in \cD \\ z<z'}} \(\s \d\cN_{[z,z']}\) = {1\o \pi}\sqrt{2\o\b}\[2\(\log N -{3\o4}\log\log N\) + \cY^{(\s)}_{N,\b}\]\qquad (\s = \pm)}
as $N\rar\i$, with random variables $\cY^{(\s)}_{N,\b} = O_\PP(1)$ expected to converge in law to limiting random variables $\cY^{(\s)}_{\i,\b}$ with tail asymptotics \eqr{piy2}. For what follows, it is useful to rewrite \eqr{Range} in unfolded form. Define an unfolded mesoscopic interval $I_\ell := [u,u+\ell]$ embedded within a larger unfolded stationary interval $ \cI_\scL \supset I_\ell$ with respective unfolded lengths $\mathscr{L}\gg\ell:= LN$. Let $\{I\}\subset I_\ell$ be the intervals over which we scan in \eqr{Range}. Then
\e{Range2unf}{\qquad \sup_{I\subseteq I_\ell}\(\s\d\cN_I\) = {1\o \pi}\sqrt{2\o\b}\[2\(\log \ell -{3\o4}\log\log \ell\) + \cY^{(\s)}_{\ell,\b}\]\qquad (\s = \pm)}
as $\ell\rar\i$ (where $\cY^{(\s)}_{\ell,\b}$ differs from $\cY^{(\s)}_{N,\b}$ by a simple $L$-dependent shift).

\ssec*{Comments}

\pg{Gaussian random model for state counts:} The imaginary field $\Phi(z)$ also enjoys a Gaussian random model \eqr{rfmp}, in exactly the same fashion as for the real field. By \eqr{counts} this model is directly inherited by the interval counts. In the unfolded language of \eqr{Range2unf}, with frequencies quantized over $\cI_\scL$, the mesoscopic interval count within $I_\ell$ obeys
\es{rfmcount}{ \d\cN_{[u,u+\ell]}&= \frac{1}{\pi\sqrt{\beta}}
\sum_{1\le k\le \mathscr{L}}{a_k\sin\({2\pi k(u+\ell)\o\mathscr{L}}\)+b_k\cos\({2\pi k(u+\ell)\o\mathscr{L}}\)\o \sqrt{k}} - (u \rar u-\ell)}
as $\ell\rar\i$, where $\EE[a_k^2] = \EE[b_k^2]=1$ and $\EE[a_kb_k]=0$. It is straightforward to compute the covariance between nearby intervals,
\e{}{\EE\[\d\cN_{[u,u+\ell]}\d\cN_{[u',u'+\ell]}\]=\frac{2}{\beta\pi^2}\sum_{1\le k\le \mathscr{L}}{1-\cos\({2\pi k\ell\o \mathscr{L}}\)\o k}\cos\({2\pi k|u-u'|\o \mathscr{L}}\)}
For $u=u'$, this is just the standard number variance $\Sigma^2(\ell)$, from which one easily recovers the bulk RMT result for mesoscopic intervals,
\e{}{\qquad \qquad\qquad\qquad\Sigma^2(\ell) \approx {2\o \b\pi^2}\log\ell \qquad (1\ll \ell \ll \mathscr{L})}
For $|u-u'|=\eps \ell$ with $\eps\neq 1$ held fixed and $1\ll \ell \ll \mathscr{L}$, the covariance is
\e{}{\EE\[\d\cN_{[u,u+\ell]}\d\cN_{[u',u'+\ell]}\]\approx \frac{1}{\beta\pi^2}\log\left|1-{1\o \eps^2}\right|}
This is the interval count analog of \eqr{EEXX2}, showing the crossover at $\eps\approx1$. 

\pg{Remark on pinned log-correlated fields:} As an aside, note that one could also consider $\sup_{z' \in \cD}\(\Phi(z')-\Phi(z)\)$, in which we have pinned $z$ as a basepoint, optimizing only over $z'$. This is the setup corresponding to so-called {\it pinned log-correlated fields}. Since $\Phi(z)$ is typically $O(\sqrt{\log N})$, and its supremum can even be much larger, this substantially modifies the law for pinned suprema: at typical scales, the leading scaling $\sim \log N$ is valid, but the subleading log-correlated term $\sim -{3\o4}\log\log N$ is already dominated by the standard deviation of $\Phi(z)$. See remarks in e.g. \c{HP,FLD,PZ22}. One instead formulates the corresponding conjecture as a statement in {\it expectation}, namely, 
\e{pinned}{\qquad \qquad\EE\[\sup_{z' \in \cD}\(\s\d\cN_{[z,z']}\)\] = {1\o \pi}\sqrt{2\o\b}\(\log N -{3\o4}\log\log N + c_{N,\b}\)\qquad (\s=\pm)}
as $N\rar\i$ for constants $c_{N,\b}$ (whose size depends on whether $\cD$ is mesoscopic or macroscopic), and with corresponding statements for the variance and higher moments. 

\pg{Remark on FHK for Riemann zeta zero counts:} It is natural to expect that the counting staircase for Riemann zeta zeros respects extremal laws of the form \eqr{Range'} and \eqr{Range}. Perhaps owing to how well-absorbed the FHK perspective has become, a conjecture for $\Im\log\z(\half+it)$, stating the {\it full} log-correlated structure, has not been explicitly stated in the literature to our knowledge -- \c{Najnudel} proved the leading order result conditional on the Riemann Hypothesis\foot{The methods of \c{ABR1,ABR2}, used to prove unconditional results for $\log|\z(\half+it)|$, would also extend to $\Im\log\z(\half+it)$ at subleading order if one is willing to assume the Riemann Hypothesis. In this regard, one essential distinction between the real and imaginary cases is that, in methods using mollified moments of Riemann zeta, phase ambiguities are invisible to absolute values of $\zeta M$, where $M$ is the mollifier; this makes the imaginary case more difficult. We thank Paul Bourgade for explaining these points.}, and \c{FLD} remark in passing that the pinned result \eqr{pinned} for C$\b$E eigenvalues ought to apply to $\Im\log\z(\half+it)$ in short intervals -- but it is clearly ``in the air.'' Let us state it now for concreteness and as a number-theoretic touchpoint for the next sections. Defining the standard zeta zero count $N(T) := \#\{\rho=\half+it: t\in(0,T),\z(\rho)=0\}$, the Riemann-von Mangoldt formula \c{IK} says that 
\e{}{N(T) = {T\o2\pi}\log{T\o2\pi e} + {7\o8} + S(T) + O\({1\o T}\)}
as $T\rar\i$, where $\pi S(T) := \arg\z(\half+iT)$ plays the role of our $\Phi$ field above. Then the $\b=2$ version of \eqr{Range}, taking the supremum for concreteness, would state the following: for $t \sim \text{Unif}([T,2T])$,
\e{fhkzetazeros}{\sup_{t\leq\t<\t'\leq t+2\pi}\d \cN^{(\z)}_{[\t,\t']} = {1\o\pi}\[2\(\log\log T - {3\o4}\log\log\log T \) + \cY_T\]}
as $T\rar\i$, where $\cY_T = O_\PP(1)$ obeys the log-correlated asymptotic \eqr{piy2}, and 
\e{}{\d \cN^{(\z)}_{[\t,\t']} = S(\t')-S(\t) = \cN^{(\z)}_{[\t,\t']} - {|\t'-\t|\o2\pi}\log \({\t\o2\pi}\)}
is the centered interval count of non-trivial Riemann zeta zeros. Unconditional proof at leading order remains outstanding. 

\sec{AdS/CFT Formulation}\label{sec4}

We are ready to adapt the FHK conjecture to the setting of AdS/CFT. Rather than adapt the original FHK formulation in terms of spectral determinants, which are subject to regularization in quantum field theory and quantum gravity even in fixed superselection sectors, we focus on interval counts, as presented in Section \ref{sec3}. In gravity, interval counts are the natural quantization of black hole entropy; in CFT, they count local operators, basic entities whose properties are central to modern rigorous approaches.

Consider a CFT in $d$ spacetime dimensions. In order to properly diagnose statistical universality in a quantum system, it is well-understood from the study of quantum chaos that one must do so within symmetry-resolved sectors \c{haake}. The global conformal group is $SO(2,d)$. A local primary operator generating an irreducible conformal multiplet is labeled by $SO(d)$ spins $\bJ := \{J_i\}$ where $i=1,\ldots,\lfloor {d\o 2}\rfloor$, and carries an $SO(2)$ conformal weight $\D$. Thus, we will analyze the spectra of local primary operators in fixed irreps of the rotation group $SO(d)$ (``fixed spin'') in the high-energy limit $\D\rar\i$.\foot{If the theory enjoys a global symmetry $G$, one should also resolve $G$, in which case $\bJ$ stands for the full set of quantum numbers under the combined rotation and global symmetry group. In $d=2$, global conformal symmetry is enhanced to two copies of a Virasoro algebra (or possibly a larger chiral algebra $\cA$), so symmetry resolution means restricting to Virasoro primary spectra of fixed spin.} 

Before defining the unfolding procedure, we give the formula for the mean spectral density at high-energy. The leading-order asymptotic spectral density for fixed spin primaries, which we denote $\rho_0^{(\bJ)}(\D)$, was determined for general $\bJ$ in \c{BLOSD}, following \c{loga, shag}. We quote the leading-order result at $\D\rar\i$ and fixed $\bJ$,
\e{asympd}{\log\rho_0^\j(\D) \approx C_d (\D+\vareps_0)\({f\o \D+\vareps_0}\)^{1\o d}\,}
where 
\e{}{C_d := {d\o d-1} \big((d-1)\text{Vol}(S^{d-1})\big)^{1\o d}\,,\quad \vareps_0 := {\sqrt{\pi}\o \G({1-d\o 2})}a_d}
The coefficient $f$ is the thermal free-energy coefficient on $S^1 \x \RR^{d-1}$, and $a_d$ is the $a$-type conformal anomaly coefficient. We emphasize that while \eqr{asympd} captures merely the leading exponentially large term of $\rho_0^\j(\D)$, which is dressed with further multiplicative loop factors \c{BLOSD}, $\rho_0^{(\bJ)}(\D)$ also includes {\it additive}, exponentially subleading terms which are nevertheless exponentially large in positive powers of $\D$: such terms are necessarily present in $d=2$ due to $SL(2,\Z)$-invariance of the torus partition function,\foot{Even in the absence of light, non-vacuum primaries, the vacuum (Cardy) contribution $|\!\sinh(2\pi P b)\sinh(2\pi P b^{-1})|^2$ to the asymptotic density includes subleading exponentials at $P,\bar P\rar\i$.} while in $d\geq 2$, they can appear as subleading saddles in thermal effective field theory \c{loga,BLOSD}.

With this in hand, we recall the standard setup for statistical sampling of deterministic spectra. Let $\cI_\cL(\D_0)$ be a stationary interval of length $\cL$ anchored at some fiducial dimension $\D_0$:
\e{}{\cI_\cL(\D_0) := [\D_0,\D_0+\cL]\,.}
The length $\cL$ is lower-bounded by the mean-level spacing, and upper-bounded by stationarity:  namely, $\rho_0^{(\bJ)}(\D_0)^{-1} \ll \cL \ll \cL_*(\D_0)$ where
\e{station}{\cL_*(\D):= |\p_\D \log\rho_0^{(\bJ)}(\D)|^{-1} \approx {d\o (d-1)C_d}\({\D+\eps_0\o f}\)^{1\o d}}
Note that $\cL_*(\D)$ is unbounded as $\D\rar\i$. Next, we define subintervals $I_L(\D) \subset \cI_\cL(\D_0)$: in unfolded coordinates $u=\rho_0^{(\bJ)}(\D) \D$, 
\es{}{I_\ell(\D) &:= [u,u+\ell]\,,\quad \quad\quad\quad\,\ell := \rho_0^{(\bJ)}(\D) L\\
I_\ell(\D)  \subset \cI_\scL(\D_0) &:= [u_0,u_0+\scL]\,,\quad\quad \scL := \rho_0^{(\bJ)}(\D_0)\cL}
with hierarchy 
\e{hierarchy}{1 \ll \ell \leq \scL \ll \scL_*(\D_0)}
where $\scL_*(\D_0)=\rho_0^{(\bJ)}(\D_0)\cL_*(\D_0)$ bounds the unfolded outer interval length. For our purposes, we further restrict $\ell \ll \scL$, such that there is an infinite set of subintervals $I_\ell(\D) \subset I_\scL(\D_0)$ as $\D\rar\i$. This leaves various possible choices for the scale $\ell$, including, as before, both macroscopic and mesoscopic cases.\foot{What follows generalizes straightforwardly to mesoscopic intervals $L \sim (\rho_0^\j(\D))^\theta$ with $\theta \in (-1,0)$, recalling the discussion in Section \ref{sec22}.} For simplicity, we choose $L = O(1)$ and $\cL \rar\i$ growing unbounded with $\D$ in a manner consistent with \eqr{station} and \eqr{hierarchy}, namely, $\cL = O(\D^\theta)$ with $\theta\in(0,{1\o d})$. 

The conjecture for extrema can now be formulated. Define the spin-$\bJ$ counting fluctuation field on spectral intervals $I$:
\e{}{\d \cN_I^{(\bJ)} = \cN_I^{(\bJ)} - \overline{\cN_I^{(\bJ)}}\,,\quad \text{where}~~\overline{\cN_I^{(\bJ)}} := \int_I d\D\,\rho_0^{(\bJ)}(\D) \approx \rho_0^{(\bJ)}(\D_0) |I|}
is the unfolded length of $I$. Then for $u \sim \text{Unif}(\cI_\scL(\D_0))$, and for every $\bJ$, 
\e{fhkcft}{\qquad \sup_{I \subseteq I_\ell(\D)} \(\s\d \cN_I^{(\bJ)}\) ={1\o\pi}\sqrt{2\o\b}\[2\(\log\ell -{3\o4}\log\log\ell\) + \mathcal{Y}_{\ell}^{(\s|\bJ)}\]\qquad (\s = \pm)}
as $\D\rar\i$, where $\mathcal{Y}_{\ell}^{(\s|\bJ)}$ are sequences of $O_\PP(1)$ random variables that converge in law to limiting random variables $\cY_{\i}^{(\s|\bJ)}$ with log-correlated tail asymptotics \eqr{piy2}. The analogous formulation holds for the amplitude law \eqr{Range'}.

The form of \eqr{fhkcft} is uniform in $\bJ$. The value of $\b$ encodes the symmetry class to which the spin-$\bJ$ spectral statistics converge. This is in turn determined by the action of CRT symmetry on the chosen $SO(d)$ irrep, and the presence or absence of antiunitary transformations that square to unity. The baseline case of neutral, real, bosonic $SO(d)$ irreps enjoys GOE statistics ($\b=1$), but in the presence of global charge or spin structures, GUE ($\b=2$) and GSE ($\b=4$) statistics are possible. 

We have adorned $\cY_\ell^{(\s|\bJ)}$ with a spin label because it could in principle depend on $\bJ$ in a manner beyond just the symmetry class: in other words, the law for $\cY_{\ell}^{(\s|\bJ)}$ may not be universal. This is the case for the Riemann zeta function, whose limiting law is thought to deviate slightly from that of the CUE characteristic polynomial. To what extent the law for $\cY_{\i}^{(\s|\bJ)}$ depends on $\b$ alone, perhaps only within a given theory, is an interesting question for future investigation. As discussed below \eqr{piy2}, a natural expectation for the tail exponent is $p=3$. 

We emphasize again that \eqr{fhkcft} is not just a bound: every typical mesoscopic short interval contains a subinterval in which the supremum and infimum achieve the log-correlated scale set by the right-hand side. In other words, it is an {\it optimal global rigidity} result. As explained in Section \ref{s23}, the leading-order term of \eqr{fhkcft} relies on quite less than the full log-correlated picture, as it follows only from Gaussian falloff of large deviations and a logarithmic number variance $\Sigma^2(\ell)$.

As for the original FHK conjecture, a milder preliminary postulate is that the interval counts $\d\cN_I^{(\bJ)}$ are log-correlated, converging as $\D\rar\i$ to a Gaussian random model of the form \eqr{rfmcount}. Writing our intervals $I$ in unfolded coordinates as $[u',u'+\ell_I]$, 
\e{rfmcountcft}{ \d\cN_I^\j= \frac{1}{\pi\sqrt{\beta}}\sum_{1\le k\le \mathscr{L}}{a_k^{(\bJ)}\sin\({2\pi k(u'+\ell_I)\o\mathscr{L}}\)+b_k^{(\bJ)}\cos\({2\pi k(u'+\ell_I)\o\mathscr{L}}\)\o \sqrt{k}} - (u' \rar u'-\ell_I)}
where $\{a_k^{(\bJ)},b_k^{(\bJ)}\}$ are i.i.d. Gaussian random variables in the limit. This is the field-level diagonalization of a mesoscopic central limit theorem with logarithmic variance, i.e. a linear ramp in the spectral form factor (see Appendix \ref{appA}). The IR frequency cutoff $k=1$ means that the interval $I_L(\D)$ with $L=O(1)$, as chosen earlier, is ``macroscopic'' with respect to the log-correlated statistics of the field $\d \cN_I^\j$. 

The law \eqr{fhkcft}, like the original FHK laws, is backed by expectations of universality of extreme log-correlated statistics \c{FB,ABB15,Arguin2017}. Stated differently, and more directly, it applies when unfolded spectral statistics at high energy converge to those of Gaussian random matrix theory -- more precisely, when the spin-$\bJ$ spectral density converges to Sine$_\b$, the limiting point process of the bulk spectrum of the Gaussian $\b$-ensemble with $\b$ determined by $\bJ$ -- on mesoscopic scales. This is not merely an analogy, but rather, an appeal to the progress since \c{FHK,FK}: if one grants the FHK conjecture for random matrices, whose core structure is now substantially proven (cf. Section \ref{sec24}), then a law of FHK form therefore holds for any deterministic point process obeying the requisite degree of convergence to random matrix statistics. Evidently, this is the case for $\log\z(\half+it)$ \c{Montgomery1973, Odlyzko1987,KSzeta,FHK,FK,ABR1,ABR2}.\foot{The Riemann zeta function is special among log-correlated processes not because it exhibits FHK extreme statistics; it is special because its extreme statistics can be computed at all. Indeed, as was emphasized in \c{ABB15,ABBRS}, certain features of the extreme value distribution of $\z(\half+it)$ were actually proven {\it before} those of random matrix characteristic polynomials! It is also the only deterministic system we are aware of among those whose local gap statistics obey random matrix universality, whose extreme gap statistics have been verified to high precision \c{BAB}: they, too, match those of random matrix theory. (See e.g. \c{blomer2016smallgapsspectrumrectangular} for an example of extreme gap statistics in a deterministic integrable system.)} An extended explanation of these points in the context of universality in quantum chaos is given in Appendix \ref{appB}.

\sec{Interpretation}\label{sec5}

There are several ways to view the postulate \eqr{fhkcft}:

\begin{enumerate}[label*=  \bf (\arabic*)]

\item As a property of large black hole microstate counts in AdS quantum gravity. 

\item As a precise and non-trivial $d$-dimensional conformal bootstrap bound on primary operator interval counts at large conformal dimension. 

\item At large $N$, as setting a precision limit for the semiclassical gravitational path integral. 

\item For $\b=1$, as a physical realization of the GOE extremal law of FHK.

\item As a hint that extremal landscapes of high-energy CFT spectra may be glassy. 

\end{enumerate}

\ni We elaborate on {\bf (1)} and {\bf (2)} below, while {\bf (3)} will be the subject of its own Section \ref{sec6}. Regarding {\bf (4)}, we note that in the analytic number theory context, the statistics of zeros high on the critical line are those of GUE, not GOE, so the GOE case is somewhat more exotic.\foot{Central zeros (those near $s=\half$) of $L$-functions may have GOE or GSE statistics when summed over infinite families labeled by conductor $q$, whose average zero density is unbounded at large $q$ \c{katzsarnak}.} Regarding {\bf (5)}, the glassiness is a stronger statement than \eqr{fhkcft}: it pertains to the structure of the spectrum beyond the local maxima, and implies the presence of a freezing transition, neither of which have we considered in detail here.

\ssec{Large black hole statistics in AdS quantum gravity}

In semiclassical general relativity, large black holes are maximally chaotic and randomly fluctuating, by several measures and in several temporal regimes (e.g. \c{Sekino,MSS,MSSYK,Cotler,SSScone,rmt2}). To the extent that microstates of the black hole Hilbert space $\mathcal{H}_{\rm BH}$ are the quantum origin of the behavior of semiclassical black holes, it is very reasonable to extend the random matrix properties of black hole states studied in \c{Cotler} to extreme regimes for mesoscopic interval counts $\cN_I^{(\bJ)}$ -- even moreso than for a generic quantum chaotic system. As we have emphasized, interval count extrema, and especially the putative $O_\PP(1)$ limiting random variables $\cY_\i^{(\s|\bJ)}$, probe large deviations not visible in marginal $k$-point correlations. 

In this sense, \eqr{fhkcft} is an extension of the spectral form factor diagnostic of random matrix behavior of large AdS black holes \c{Cotler, SSScone}: it is a statement of universality of extreme value laws for log-correlated fields. Even before reaching the extremal regime, the assertion is that large black hole microstate counts exhibit log-correlated statistics, admitting a random Fourier model of the type \eqr{rfmcountcft} to leading order in the high-energy limit. 

This central theme of this work raises the obvious question: if state counts in AdS/CFT exhibit log-correlated statistics, {\it what is the log-correlated field}? In quantum gravity terms, the answer is just the log-spectral determinant over spin-$\bJ$ black hole microstates\foot{This requires regularization to define rigorously due to the infinite number of high-energy states, even at finite $N$; this is one reason to favor interval counts, which regularize by way of taking a ratio.}
\e{Phidef}{\Phi^{(\bJ)}\!(E) = \log \det(E-H_{\rm BH}^{(\bJ)}) - \EE\[\log \det(E-H_{\rm BH}^{(\bJ)})\]}
with logarithmic autocorrelation
\e{PhiPhi}{\quad \EE\[X(E)X(E')\] \sim -C_X\log|E-E'|\,,\quad X \in \{\Re[\Phi^{(\bJ)}],\Im[\Phi^{(\bJ)}]\}}
as $E,E'\rar\i$ in AdS units for some $C_X >0$. In general, $\Phi$ need not have a bulk brane interpretation: although it does in JT gravity, this is a coincidence special to the AdS$_2$ setting, and there is nothing inherently low-dimensional about $\Phi$. We discuss this further in the large $N$ section. 

What is $\cY_\i^{(\s|\bJ)}$ in AdS/CFT? To what extent does it depend on $\bJ$ beyond the determination of $\b$? Does it depend on physical couplings? Certainly the {\it average} count $\overline{\cN_I^{(\bJ)}}$ depends on couplings, because the classical black hole entropy does, via the Wald entropy formula. But the random extremal fluctuations $\cY_\i^{(\s|\bJ)}$ could plausibly be {\it universally} those of a Gaussian random matrix ensemble (perhaps with a deterministic shift), and thus characterized by the randomly-shifted Gumbel of \c{PZ22}. 

\ssec{State-counting in CFT and the bootstrap}

In CFT terms, \eqr{fhkcft} is a statement of universal log-correlated extremal statistics of spin-$\bJ$ local primary operators at large conformal dimension. The log-correlated field is $\Phi^\j(\D)$, dual to the field defined in \eqr{Phidef}, where $H^\j_{\rm BH}$ is dual to the spin-$\bJ$ Hamiltonian of the CFT on $S^1 \x \RR^{d-1}$. The CFT spectra in question admit a random Fourier model of the type \eqr{rfmcountcft} with Gaussian $\{a_k,b_k\}$ to leading order in the $\D\rar\i$ limit. This model is an extension of a mesoscopic CLT with a Gaussian random matrix form factor (cf. Appendix \ref{appA}).\foot{We expect that an OPE analog of \eqr{fhkcft} may hold, with a log-correlated structure for OPE-weighted counting statistics at $\D\rar\i$, as supported by a linear ramp of a generalized spectral form factor with operator insertions \c{Belin:2021ibv}. The role of random matrix theory would be played by the OPE Randomness Hypothesis \c{ORH}.}

Viewed as a bootstrap-type bound, \eqr{fhkcft} is extremely stringent. Conceptually speaking, {\it it is a statistical version of a Tauberian formula.} In the conformal bootstrap setting, bounds on CFT data integrated over high-energy intervals fall under the purview of Tauberian analysis, whose purpose is to make asymptotic formulas rigorous. In the statistical setting, the relevant analog of Tauberian analysis is the purview of extreme value theory! In other words: given a spectral statistical universality class $\sfU$, what it means to bound a spectral field $X_\sfU$ in the sense of the bootstrap is to constrain large excursions of $X_\sfU$ subject to the probabilistic laws of $\sfU$.  This is exactly the domain of extreme value theory. 

In the present case, with $X_\sfU \mapsto \d \cN_I^\j$ and $\sfU \mapsto$ Sine$_\b$, the FHK log-correlated universality makes a sharp statistical prediction. Even the robust leading-order result
\e{bootlead}{\qquad \sup_{I\subseteq I_\ell(\D)} \(\s\d \cN^\j_{I}\)  \stackrel{\PP}{\longrightarrow} {2\o\pi}\sqrt{2\o\b}\log\ell \qquad (\s=\pm)}
proven in \c{HP, CFLW, BLZ} is stronger in magnitude than any current results from the $d$-dimensional bootstrap:

In $d>2$, while \c{espin,RychkovYvernay2016,QiaoRychkov2018,MZ19,VanRees2025} developed Tauberian methods for OPE convergence inside four-point functions, there are no such bootstrap bounds on purely spectral data. It would be very interesting to test \eqr{bootlead} numerically in explicit CFTs, such as the 3d Ising model. Depending on the rate of approach to semiclassics, the asymptotic behavior could already be visible above some dimension $\D_0$ which is accessible numerically. It should also be a priority to verify the logarithmic number variance $\Sigma^2(\ell) \propto \log\ell$, a simpler target that has not yet been tested in the 3d Ising model.

In $d=2$, rigorous bounds on interval counts $\cN^{(J)}_I$ of Virasoro primaries {\it have} been derived, by combining modularity and unitarity with analytic extremization methods \c{baursasha, MP}. The sharpest such results may be found in \c{MP}, which we quote here in our notation. For intervals $I$ centered at $\D$ with unfolded length
\e{}{\ell_I^{(J)} := \rho_0^{(J)}(\D)|I|\,,}
where $\rho_0^{(J)}(\D)$ is the spin-$J$ (Cardy) density for Virasoro primaries, 
\e{mpbound}{\half-|I|^{-1} \leq \liminf_{\D\rar\i} {\cN^{(J)}_I\o \ell_I^{(J)}} \leq \limsup_{\D\rar\i} {\cN^{(J)}_I\o \ell_I^{(J)}} \leq {\half+|I|^{-1}}}
for all $|I|>0$. The same formula holds for counts not graded by spin, but with $\half \rar 1$. The $\half$ is a symptom of the generality of this result: it applies to all CFTs obeying modular invariance and unitarity, including holomorphically-factorized CFTs. The latter, by invariance under the $T$ transformation of $SL(2,\ZZ)$, must have integer-quantized left- and right-moving conformal weights 
\e{}{h={\D+J\o 2}\,,\quad\hb = {\D-J\o 2}\,,}
and, therefore, a dimension gap of exactly $\D-\D'=2$ between consecutive primaries of fixed spin. This explains the $\half$ in the lower bound, and the method of computation \c{MP} maps it to the upper bound.

Now, how do these bounds \eqr{mpbound} change when we exclude rational CFTs? In non-factorized CFTs, the modular obstruction $h,\hb\in\Z$ disappears, so even the fixed-spin bounds should have $\half \rar 1$. That's not terribly interesting. More interesting is what happens upon imposing non-trivial spectral statistics. A general folk expectation is that in any interacting CFT whose high-dimension spectrum is in some sense chaotic, the lower and upper bounds should converge up to small errors, i.e.
\e{e0}{\liminf_{\D\rar\i} {\cN^{(J)}_I\o \ell_I^{(J)}} \approx \limsup_{\D\rar\i}  {\cN^{(J)}_I\o \ell_I^{(J)}} \approx 1}
over any mesoscopic stationary interval $I$. A weaker, and slightly more precise, form of this is a sup norm bound:
\e{}{\sup_{I\subseteq I_\ell(\D)} \left|\d \cN^{(J)}_{I}\right| = o(\ell)\,,}
as $\D\rar\i$, where we recall that $\overline{\cN^{(J)}_I} = \ell_I^{(J)}$ and hence $\d \cN^{(J)}_{I} = \cN^{(J)}_{I}-\ell_I^{(J)}$. But this is precisely what follows from the leading-order probabilistic equality \eqr{bootlead} which further resolves the $o(\ell)$ into a logarithm. So a sharper form of \eqr{e0} is indeed true: in a $c>1$ Virasoro CFT whose fixed-spin primary spectral statistics converge to those of bulk random matrix theory over mesoscopic scales as $\D\rar\i$, \eqr{bootlead} holds. (By the argument in \c{Yan}, in $d=2$ one has GOE statistics, $\b=1$, for any $J$.) 

Let us place this in a broader context. An outstanding goal for the next-generation bootstrap of $d$-dimensional CFTs has been to reconceptualize the bootstrap in a manner that incorporates high-energy spectral statistics. The most physically direct interpretation of this is to use the statistics {\it themselves} to organize the space of CFTs. This is akin to how families of $L$-functions are defined in number theory \c{katzsarnak}. For CFTs with random matrix statistics in the precise sense defined earlier, we are stating a bootstrap result \eqr{bootlead} for their high-energy state counts, implied by FHK. An alternative approach would be to bootstrap a subspace of CFTs subject to extra deterministic assumptions that are designed to exclude theories with extra structure and/or possible formal solutions with no physical realization. This method, in some sense a ``purist'' approach, can be difficult to motivate convincingly: one can question the physical meaning and applicability of the formal assumptions involved. Hybrid approaches are also possible, in which the assumptions are themselves statistical but milder in nature than the output. Altogether, it would be very interesting to ask what independent extra restrictions on CFT data would lead to a complementary proof of a deterministic analog of \eqr{bootlead}.\foot{A constructive approach to bootstrap in 2d CFT incorporating similar ideas, applied to deriving a rigorous upper bound on the primary spectral gap, will be introduced in \c{nmb}.} 

Needless to say, in any dimension $d$ the subleading term of \eqr{fhkcft} and the random variables $\cY_\i^{(\s|\bJ)}$ are much more refined properties of CFT data, and would be extremely interesting to understand further: they characterize the genuine log-correlated structure of the spectral determinant $\Phi^\j(\D)$. Ultimately, the picture we are suggesting is that asymptotic spectra in AdS/CFT form a glassy log-correlated extremal landscape. 

\sec{Large $N$ and Semiclassical AdS/CFT}\label{sec6}

At large $N$, the law \eqr{fhkcft} has interesting implications for semiclassical gravity and AdS/CFT. By ``large $N$'' we refer to the asymptotic behavior of a sequence of CFTs $\{\mathcal{T}_N\}$ labeled by $N\in \ZZ_+$, such that a measure of the number of degrees of freedom, call it $c_N$, diverges as $N\rar\i$:
\e{}{\{\mathcal{T}_N\}_{N\in\Z_+}~~\text{s.t.}~~ \lim_{N\rar\i} c_N = \i}
In this section we mostly suppress the $\bJ$ label for clarity, though every entropy relation, interval count, spectral density, etc. holds separately for all fixed $\bJ$. 

Let us rewrite \eqr{fhkcft} in gravity language. The first observation is that the mean density $\rho_0(\D)$ is canonically identified with the smooth part of the spectral density extracted from the one-boundary Euclidean semiclassical gravitational path integral (GPI). Correspondingly, the mean interval count $\overline{\cN_I} = \rho_0(\D) |I|$ is identified with the smooth semiclassical gravity result, which we call $\cN_I^{\rm (grav)}$: 
\e{gravav}{\cN_I^{\rm (grav)} := \overline{\cN_I}}
So in semiclassical AdS/CFT, \eqr{fhkcft} determines the scale of fluctuations of high-energy microstate counts around the coarse-grained bulk result: 
\e{fhkgrav}{\sup_{I \subseteq I_\ell(\D)} \(\cN_I -  \cN_I^{\rm (grav)} \) ={1\o\pi}\sqrt{2\o\b}\[2\(\log\ell -{3\o4}\log \log\ell\) + \mathcal{Y}_{\ell}\]}
as $\D\rar\i$ (for every $\bJ$), and likewise for the infimum. Let us rewrite this in terms of semiclassical entropy, in order to develop gravity intuition. Define an entropy $S_0$ (for every $\bJ$) from the leading-order density as
\e{}{\log \rho_0(\D) := S_0(\D) + O(\log S_0(\D))}
Then recalling that $\ell = \rho_0(\D) L$, where $L$ is upper-bounded by the stationarity constraint \eqr{station} to be a power of $\log S_0$, the leading-order term $\log\ell$ is linear in $S_0$. The terms in \eqr{fhkgrav} thus have respective scalings 
\e{scaling}{\cN_I^{(\rm grav)} \sim e^{S_0}\,,\quad \log\ell \sim S_0\,,\quad \cY_\ell = O_\PP(1)}
We are taking $\D\gg c_N \gg 1$, but later we mention other regimes. 

The relation \eqr{fhkgrav} is very rich, not least because it identifies an erratic behavior $\cY_\i$ in the large $N$ microcanonical spectral data. This is a limiting random variable for spectral fluctuations around the mean semiclassical gravity result. It is not a correction to the smooth density, computed by a semiclassical bulk saddle: it is a statistic of the erratic residual. In gravitational terms, this is evocative of the ``half-wormhole'' concept of \c{sssy}. It would seem reasonable to interpret the random variable $\cY_\i$ as indicating the absence of an asymptotic series in the sum over topologies in semiclassical gravity with a single asymptotic boundary.

The fact that \eqr{fhkgrav} is an equality -- not just a one-sided bound -- has interesting implications in view of the entropic scalings \eqr{scaling}, which we explain in turn below:

\begin{enumerate}[label*= (\sf \roman*)]

\item Viewed as a lower bound, it sets a parametric {\it precision limit} of $O(e^{-S_0})$ for the semiclassical gravitational path integral. 

\item Viewed as an upper bound, it implies a certain semiclassical {\it completeness} relation for exponentially large contributions to the density of states.  

\end{enumerate}

\pg{Precision limit:} First, the lower bound. \eqr{fhkgrav} says that the amplitude of extremal counting fluctuations within mesoscopic intervals $I \subset I_\ell(\D)$ is {\it no smaller} than the right-hand side. In entropic terms \eqr{scaling}, this is a suppression of $O(S_0e^{-S_0})$ relative to the mean. Even if the deterministic centering term could somehow or sometimes be understood as a semiclassical bulk effect, the random fluctuations of $\cY_\i$, which are $O_\PP(1)$ and hence suppressed by $O(e^{-S_0})$, set an irreducible stochastic baseline. In other words, {\it in the computation of a microcanonical density of spinning black hole states, the semiclassical GPI cannot resolve effects down by more than $e^{-S_0}$ relative to the leading saddle, due to order one erratic fluctuations in the spectrum.} 

This is a sharpening of the common heuristic that in the computation of generic gravity observables (absent enhanced structures, such as supersymmetry), semiclassical geometry cannot capture all non-perturbative effects in $G_N$. It is a sharpening because it identifies a specific observable ($\d \cN_I^\j$), a specific functional of this observable (its extremal excursions), and the shape of the obstruction (the log-correlated law \eqr{fhkgrav}). Note that the exponent is $D$-independent, despite the absence of a genus expansion in AdS$_{D>2}$. 

Our interpretation here relies on the standard identification \eqr{gravav}. In AdS/CFT, geometric saddle points with a single asymptotic boundary are regarded as contributing, after inverse Laplace transform, to a smooth density of states in the boundary theory, where the word ``smooth'' is meant in the coarse-grained sense of the quantum chaos literature: there is an asymptotic density which is non-fluctuating over mesoscopic scales. It is certainly true at leading order in $\D$ from the standard AdS/CFT identification of the leading-order contribution to $\log \rho_0(\D)$ with semiclassical black hole entropy; the semiclassical GPI also contains (usually infinitely) many subleading smooth contributions, including from sums over subleading geometric saddles and, possibly, other non-saddle contributions (e.g. smooth off-shell geometries). The identification \eqr{gravav} is merely the modest statement that the smooth density is included among the bulk contributions. Indeed, the law \eqr{fhkgrav} is to be read as a statement about what {\it else} the bulk includes. This takes us to the next point. 

\pg{Completeness:} Now viewing \eqr{fhkgrav} as an upper bound, the amplitude of extremal counting fluctuations around the mean over typical intervals cannot be {\it larger} than $O(S_0e^{-S_0})$ relative to the leading saddle: {\it all exponentially large contributions $\sim e^{\a S_0}$ with $\a >0$ must be included in $\cN_I^{(\rm grav)}$.} This is the converse of the precision limit: for every $\a>0$ term present in the asymptotic expansion of state counts $\cN_I$, the mean term $\cN_I^{(\rm grav)}$ must include it, else the extremal equality \eqr{fhkgrav} would be violated.

This leads us to a brief tangent about subleading geometries in gravity. In AdS/CFT, sums over semiclassical saddles are not known to account for all exponentially large terms in the microcanonical density of states. Moreover, the correct contour for the GPI may also include off-shell geometries, even contributing to one-boundary quantities, and indeed, there is evidence that this is so \c{MW,SSScone,CJ,sssy}. This leads us to raise a question: are {\it all} $\a>0$ terms accounted for by saddles? Conversely, 
\e{offq}{\exists\, \a>0 ~~\text{s.t.}~~\rho_{\text{off-shell}}(\D) \sim e^{\a S_0(\D)}\,?}
By $\rho_{\text{off-shell}}(\D)$ we mean the contribution from the sum over all one-boundary off-shell geometries. The novel relation of this question to extremal statistics is that \eqr{fhkgrav} relates the {\it size} of off-shell amplitudes to their inclusion or exclusion in the semiclassical GPI. If the answer to \eqr{offq} is ``yes,'' then such an off-shell term must be included in $\cN_I^{(\rm grav)}$. If the answer to \eqr{offq} is ``no,'' then gravity obeys what we might call a {\it saddle completeness} relation: namely, that semiclassical saddles account for all exponentially large terms, with $\a>0$. In view of the precision limit of $O(e^{-{S_0}})$ set by the erratic statistical fluctuations, saddle completeness would thus imply that off-shell geometries cannot be operationally resolved in the semiclassical GPI. 

Experimental evidence in low-dimensional examples of holography, while rather limited, augurs well for saddle completeness: in all controlled off-shell computations of which we are aware, $\a\leq 0$. In particular, this is the case for the Maxfield-Turiaci amplitudes, computed in JT gravity with defects but regarded as predictions for Seifert manifold amplitudes in AdS$_3$/CFT$_2$ \c{MT}; for certain AdS$_3$ off-shell amplitudes computed as slight deformations away from Virasoro TQFT \c{boris,deBoer:2025rct}; and for the expected scaling of CFT$_2$ matrix-tensor model amplitudes \c{belinetal,Jafferis:2025vyp}. In the Maxfield-Turiaci case, a short calculation gives\foot{This is strictly valid for large $J$, but it is believed that the scaling holds for large $cJ$, hence for all $|J|>0$.} 
\e{}{\a = 1-\sum_{a=1}^k \(1-q_a^{-1}\)}
where the integers $k$ and $\{q_a\}$ characterize the number of defects on the base disk and their respective opening angles $\phi_a = {2\pi/ q_a}$. Off-shell geometries obey $k\geq 2$ and $q_a \geq 2$ \c{MT}. Then $\a$ is maximized by $k=q_1=q_2=2$, which gives $\a=0$. In the deformed Virasoro TQFT amplitudes, the density is proportional to an expansion parameter
\e{}{e^{-{c\o 6\pi}\text{Vol}(\cM)}}
where $\text{Vol}(\cM)>0$ is the volume of a hyperbolic 3-manifold, and moreover has $\D$-dependence suppressed relative to Cardy at $\D\rar\i$. Due to the volume conjecture, any amplitude constructed by gluing AdS$_3$ trumpets (essentially a cutoff version of a $T^2 \x I$ wormhole \c{CJ}) to Virasoro TQFT amplitudes with boundary is suppressed by this factor \c{vtqft}. It would be interesting to seek direct arguments for or against saddle completeness, in AdS$_3$ or higher dimensions. 

\pg{Relation to JT gravity and universe field theory:} In JT/RMT duality \c{SSS}, the ``brane operator'' $\psi(E) := \exp(\Phi(E)) = \det (E-H)$ (a probe analog of eigenbranes \c{eigen}) is dual to a so-called FZZT brane with an AdS$_2$ bulk spacetime interpretation as a generating function for universe field theory \c{uft}. In higher dimensions, these spectral branes are not expected to have a spacetime brane interpretation. The reason is that spacetime branes live in the space of theories, while spectral branes live in the space of eigenvalues: the two are equivalent only when the theory is a theory of eigenvalues -- precisely as in JT/RMT duality.\foot{In other words, a single $D$-brane takes $N \rar N-1$, while an eigenbrane takes $N_E \rar N_E-1$, which differ unless $N := N_E$.} In AdS$_{D>2}$ holography, the number of degrees of freedom at energies above the large black hole threshold is instead exponential in a power of $N$. 

So what generalizes to AdS$_{D>2}$ from the JT/RMT and universe field theory picture is not a spacetime worldvolume boundary condition: it is the spectral object $\Phi(E)$ and its universal log-correlated statistics \eqr{PhiPhi}, in each symmetry-resolved sector as $E\rar\i$. 

The log-correlated two-point function \eqr{PhiPhi} of $\Phi(E)$ was observed in the original paper \c{SSS}, without extension to the full slate of expected universal statistics. As it is a clear priority vis-\`a-vis \eqr{fhkgrav} to understand $\cY_\i$ from a bulk point of view in AdS$_D$ in general $D$, doing so in JT gravity would be a logical first step. In fact, it would be of value in its own right to extend the JT/RMT dictionary to the level of the original FHK conjecture for extremal statistics of random matrix characteristic polynomials: an initial goal should be to ratify the original FHK laws \eqr{fhk1} and \eqr{fhk2global} in JT gravity language and, perhaps (optimistically), to prove the log-correlated tail \eqr{piy} (yet to be fully proven probabilistically). This would set the stage for probing the law \eqr{fhkgrav} for interval counts in higher dimensions, and for determining the exponent $p$ in the tail asymptotic \eqr{piy2} of the random variable $\cY_{\i}$. In doing so, it would be valuable to formulate a bulk wormhole dual of the moments of $\cY_\i$, or the ``moments of moments'' of \c{BaileyKeating2019Moments}.

\ssec{On erratic $N$-dependence of high-energy spectra}\label{sec61}

In our discussion, two types of randomness have naturally emerged: first, the limiting random variables $\cY_\i$ characterizing the extreme fluctuations of $\d\cN_I$ in \eqr{fhkgrav}; and second, the Gaussian random models for state counts $\d \cN_I$ in \eqr{rfmcountcft}. It is natural to ask how these interface with recent discussions on erraticness in the $N$-dependence of holographic CFT data \c{Schlenker:2022dyo,Liuclosed, Gesteau, KFWpre, Liu, Kling, KFW}.

Very briefly, the observation being addressed in these works, based on properties of wormhole amplitudes in gravity, is that in the approach to large $N$ of the sequences $\{\mathcal{T}_N\}$, spectral data above the large black hole threshold seemingly must behave erratically as a function of $N$. This should be distinguished from, but is nevertheless similar to, the more familiar notions of randomness at high energy for chaotic spectra in a fixed theory. Indeed, one may characterize the overall spirit of these works as asking to what extent theory space may be treated statistically, in some analogy to spectral space.

The randomness of the limiting random variables $\cY_\i$ is essentially in $\D$, not in $N$. One can introduce $N$-dependence by taking the interval $I_\ell(\D)$ to be anchored at $\D \propto c_N$, say, such that the asymptotic limit that defines \eqr{fhkcft} is instead taken as a function of $N$: the convergence in law to FHK extreme statistics is now {\it in $N$}, rather than in $\D$ per se. This would be called a ``family aspect'' limit in the number theory context. While useful, this is not yet directly addressing the erratic $N$-dependence puzzles summarized above. 

On the other hand, the Gaussian random model \eqr{rfmcountcft} for the counting fluctuation $\d \cN_I$ is exactly the kind of object we should consider. Indeed, if $\d \cN_I$ converges in law to a Gaussian random variable, it cannot admit an asymptotic $1/N$ expansion. This is a good starting point for applying the perspective of \c{Liu, KFW} directly to CFT data: in particular, we can characterize erraticness criteria of \c{KFW} in terms native to AdS/CFT. 

For the purposes of the present computation, we simplify notation relative to \eqr{rfmcountcft} and consider a random model for an observable $X_N(z)$ of the general form
\e{}{X_N(z) = \sum_{k \leq K_N}{a_k(N)\psi_k(z)\o \sqrt{k}}}
with eigenfunctions $\psi_k(z)$ and frequency cutoff $K_N$, and asymptotically Gaussian coefficients
\e{}{\EE[a_k(N)a_\ell(N)] \approx \d_{k\ell}}
as $N\rar\i$. The field $X_N(z)$ may stand for $\d \cN_I$, or another observable which is asymptotically normally distributed. Our question is: {\it how are the $\{a_k(N)\}$ constrained by the erraticness criteria of \c{KFW}?}

The answer is that erraticness in $N$ is not just about asymptotic Gaussianity of $\{a_k(N)\}$: it is about their {\it joint law in $N$}, and in particular, the presence of {\it decorrelation in $N$}. What matters is how the entire collection of modes reshuffles as $N$ grows. 

We can consider the mode covariance in $N$ for fixed $k$, with a diagonal ansatz
\e{}{\EE[a_k(N)a_\ell(N+\d N)] \approx r_k(N;\d N)\d_{k\ell}}
as $N\rar\i$, with expectation value defined by sampling over a matched asymptotic high-energy interval. This gives the field covariance at a fixed point $z$, call it $R_N(\d N)$,
\e{Rrrel}{R_N(\d N) := \EE[X_N(z)X_{N+\d N}(z)] \approx \sum_{k\leq K_N} {r_k(N;\d N)\o k}}
where we have taken $|\psi_k(z)|^2=1$ for simplicity. Henceforth we implicitly normalize by $R_N(0) \approx \log K_N$. We distinguish two broad classes of behavior for the modes $\{a_k(N)\}$:

\begin{itemize}

\item If the modes are ``nested'' in $N$, such that $a_k(N) \approx a_k(N+\d N)$ for $\d N>0$, then the field will be highly correlated in $N$ as one approaches the asymptotic Gaussian model. 

\item On the other hand, $r_k(N;\d N)$ can {\it decay} in $N$ for sufficiently large separations $\d N$. 

\end{itemize}
\noindent The latter is the signature of erraticness in $N$ that is relevant for AdS/CFT wormhole physics. \c{KFW} identified a certain decorrelation condition apparently implied by gravity domain wall computations (also performed in \c{cj22}) interpolating from $\mathcal{T}_N \rar \mathcal{T}_{N+1}$. In terms of the covariance above, the condition is an onset of effective independence once $\d N$ exceeds a certain correlation length $\xi_N$: that is, there exists a scale $\xi_N$ such that $R_N(\d N \gg \xi_N) \ll 1$. A more specific falloff condition is implied by the double-cone computation, namely, an exponential falloff 
\e{kfweq1}{\qquad \qquad\qquad R_N(\d N \gg \xi_N) \lesssim e^{-\d N/\xi_N}\qquad (\d N \gg \xi_N)}
Moreover, if the Mellin averaging of \c{KFW} is to reproduce $1/N$ expansions, the correlation length itself should decay superpolynomially in $N$,
\e{kfweq2}{\xi_N \ll N^{-B}~\forall\,B>0}
Given \eqr{Rrrel}, one can write sufficient conditions on the modes $\{a_k(N)\}$ for these to hold -- for instance, a superpolynomial decorrelation for $\d N \gg \xi_N$, for every $k$ as $N\rar\i$,
\e{modecond}{\exists\,\xi_N \ll N^{-B}~\forall\,B>0~~\text{s.t.}~~\sup_k|r_k(N;\d N)| \lesssim e^{-\d N/\xi_N}~~\forall \,\d N \gg \xi_N\,.}
The modes may remain correlated for small $\d N$. In the physical setting of AdS/CFT, in which the index $N\in\ZZ_+$ is defined a priori as an integer label for a family of CFT data, $\d N \in \ZZ_+$ as well. If $c_N$ scales as a power of $N$, as in canonical examples, the characteristic scale of these effects is super-exponentially small in any positive power of the central charge. (This condition \eqr{modecond} is sufficient, but certainly stronger than necessary, for \eqr{kfweq1} to hold.)

Summarizing, this discussion provides an embedding of the criteria of \c{KFW} into Gaussian random models for state counts $\d \cN_I$, whose applicability in CFT has been motivated elsewhere in this work. In particular, our remarks are intended to complement the trigonometric toy models of \c{KFW}, by articulating the required behavior of random models of high-energy CFT data that naturally emerge from central limit theorems. 

\sec*{Acknowledgments}

We are grateful to Paul Bourgade, Jon Keating, Jonah Kudler-Flam, Ji Hoon Lee, Julian Sonner, Edward Witten, Jakub Zakrzewski and Yanjun Zhou for helpful discussions, and to Paul Bourgade and Jon Keating for comments on a draft. We thank the Isaac Newton Institute for Mathematical Sciences, Cambridge during the workshop ``Physics and automorphic $L$-functions: gravity, conformal field theory and number theory,'' and the Galileo Galilei Institute for Theoretical Physics during the conference ``Pathways to Quantum Black Holes: from Effective Theories to Exact Methods,'' for hospitality and inspiration for this work. 
 
\begin{appendix}

\sec{Random Fourier model}\label{appA}
Here we explain how the random Fourier model is derived by formal diagonalization of a mesoscopic central limit theorem (CLT) for linear statistics. 

We work in unfolded coordinates $u$. Let $X(u)$ be a linear statistic of an underlying centered spectral density $\d\rho$,
\e{}{X(u) = \int_\RR dv f_u(v) \d\rho(v)}
for some test function $f_u(v)$. The CLT determines the covariance as a quadratic form in the test function, with kernel given by the connected pair correlation of the density fluctuation: in frequency space,
\e{}{\EE[X(u)X(u')] = \int_\RR dk\, m_\rho(k) \hat f_u(k) \overline{\hat f_{u'}(k)}}
where 
\e{}{m_\rho(k) = \int_\RR dv \,e^{-2\pi i k v}\, \EE[\d\rho(v)\d\rho(0)]}
is the Fourier-transformed connected pair correlator (assuming stationarity). $m_\rho(k)$ is otherwise known as the form factor, or structure factor, of $\d\rho$; in high-energy parlance, it is nothing but the {\it microcanonical spectral form factor}, usually denoted $K_E(\t)$, upon identifying $k :=\t$. For a deterministic spectrum, $\EE[\cdot]$ is defined by a statistical sampling over mesoscopic windows, as described in the main text. 

Formally, one may diagonalize the covariance by representing $X(u)$ as
\e{Xdiag}{X(u) = \int_\RR dk \sqrt{m_\rho(k)} \hat f_u(k) dW(k)}
where $dW(k) = \overline{dW(-k)}$ is complex Gaussian white noise.  For a CLT with logarithmic variance, the form factor has a linear ramp $m_\rho(k) \propto |k|$ at frequencies $|k| \lesssim O(1)$. This is famously the case for random matrix theory, with ramp coefficient $2/\b$. Discretizing frequencies over a stationary interval of unfolded length $\mathscr{L}$ gives a random model
\e{}{X(u) = \sqrt{2\o\b}\[\Re\sum_{|k| \leq \mathscr{L}} g_k \sqrt{|k|}  \hat f_u(k)\]}
where $g_k$ are complex Gaussian variables obeying $\EE[g_k\overline g_{k'}] = \d_{kk'}$. 
Thus we see that, granting \eqr{Xdiag}, the linear statistic $X(u)$ will have a log-correlated covariance if the test function $\hat f_u(k)$ has a $1/|k|$ divergence at small $|k|$. This is precisely the case for a log-determinant or an interval count, two examples relevant for the main text. Note that macroscopic (as opposed to mesoscopic) statistics in Gaussian ensembles are sensitive to the shape of the semicircle eigenvalue distribution, and the Fourier eigenbasis generalizes accordingly in a manner respecting the symmetries (e.g. \c{FyodorovSimm2016}).

\sec{Universality}\label{appB}

Here we make a few extra comments about the FHK conjectures in the context of random matrix universality.

The original two FHK conjectures \eqr{fhk1} and \eqr{fhkzeta} made essentially distinct assertions. The first was that the log-characteristic polynomial of random matrices has the extreme value statistics of a Gaussian log-correlated field, and moreover, that it exhibits the statistical mechanical properties of random energy landscapes. The second was discussed at the end of Section \ref{s23}: namely, $\log |\z(\half+it)|$ has the extreme value statistics of the log-characteristic polynomial of $N\x N$ GUE random matrices with $N\approx \log{t\o2\pi}$, and therefore, its extremes also lie in the log-correlated universality class. This extended previous empirical relations between $\z(\half+it)$ and GUE random matrices to a conjectural matching of extreme statistics. (Alternatively, \eqr{fhkzeta} may be asserted by directly appealing to universality of log-correlated extremes.)

A natural question is how strong this ``extremal extension'' is in the context of quantum chaos. For deterministic spectra believed to obey random matrix statistics, the BGS conjecture \c{BGS} is normally understood, after appropriate unfolding, as a universal convergence of local spectral correlations in a semiclassical limit. However, it must be emphasized that the intended universality is not usually stated very precisely, specifically at the level of the {\it limits} of the universality when it applies: there is no canonical statement of how strong the aforementioned convergence is supposed to be. In practice, what is usually observed empirically is local point process convergence to Sine$_\b$ on microscopic or mesoscopic scales, and in turn, the universality of all $k$-point fluctuation measures.\foot{One may phrase this formally as something like the following: for a point process $X(\g)$, for every local observable $F(X(\g))$ that remains bounded within some stationary interval $I$, its average over $I$ converges to the same quantity in Sine$_\b$.} This mesoscopic ``point process universality'' does not a priori guarantee extreme value universality, which requires control over large-scale statistics. Nevertheless, in deterministic examples in which computations are possible and random matrix universality holds locally, it {\it does} extend to extreme values: in such cases, log-correlated universality is affirmed even at the level of extreme value statistics. 

Now, the impressive progress since \c{FHK,FK} has proven the FHK law for Gaussian random matrices up to the specific form of the limiting random variable. Given this, one can logically rephrase the Riemann zeta FHK law \eqr{fhkzeta} as just one example of a more general principle: an FHK law holds for any deterministic point process obeying sufficiently strong convergence to random matrix statistics. More sharply stated, FHK-type laws are expected when convergence to Sine$_\b$ holds on mesoscopic scales, such that the log-covariance holds on unfolded frequency scales $\ell^{-1} \ll |k| \lesssim 1$ (i.e. in high-energy parlance, until the ramp-plateau transition). What singles out the Riemann zeta function is, therefore, not that an FHK law applies, but that, as in various other contexts, it provides a tractable case study in which this can be, and has been \c{ABR1,ABR2}, independently verified.

As with all statements of universality for ergodic systems, there can in principle be exceptions to the universality of FHK laws for log-correlated systems, due to scar-like effects. Since the systems in question are taken to have local random matrix statistics, these eigenvalue scars must be even more finely-tuned than in other contexts, ``hiding'' in the extrema without spoiling coarser observables like sine kernel pair correlation. We are not aware of any ensemble or chaotic Hamiltonian known to exhibit this hypothetical behavior.

\end{appendix}

\end{spacing}

\bibliographystyle{JHEP}
\bibliography{fhk_zeta_random_matrix_test_v2}

\providecommand{\href}[2]{#2}\begingroup\raggedright\begin{thebibliography}{10}

\bibitem{FHK}
Y.~V. Fyodorov, G.~A. Hiary, and J.~P. Keating, {\it {Freezing Transition,
  Characteristic Polynomials of Random Matrices, and the Riemann
  Zeta-Function}},  {\em Phys. Rev. Lett.} {\bf 108} (2012), no.~17 170601,
  [\href{http://arxiv.org/abs/1202.4713}{{\tt arXiv:1202.4713}}].

\bibitem{FK}
Y.~V. Fyodorov and J.~P. Keating, {\it {Freezing Transitions and Extreme
  Values: Random Matrix Theory, {$\zeta(1/2+it)$}, and Disordered Landscapes}},
   {\em Philos. Trans. Roy. Soc. A} {\bf 372} (2014), no.~2007 20120503,
  [\href{http://arxiv.org/abs/1211.6063}{{\tt arXiv:1211.6063}}].

\bibitem{Montgomery1973}
H.~L. Montgomery, {\it {The Pair Correlation of Zeros of the Zeta Function}},
  {\em {Proc. Amer. Math. Soc.}} {\bf 24} (1973), no.~3 519--522.

\bibitem{Odlyzko1987}
A.~M. Odlyzko, {\it {On the Distribution of Spacings Between Zeros of the Zeta
  Function}},  {\em {Mathematics of Computation}} {\bf 48} (1987), no.~177
  273--308.

\bibitem{Cardy:1986ie}
J.~L. Cardy, {\it {Operator Content of Two-Dimensional Conformally Invariant
  Theories}},  {\em Nucl. Phys. B} {\bf 270} (1986) 186--204.

\bibitem{loga}
S.~Bhattacharyya, S.~Lahiri, R.~Loganayagam, and S.~Minwalla, {\it {Large
  Rotating {AdS} Black Holes from Fluid Mechanics}},  {\em JHEP} {\bf 09}
  (2008) 054, [\href{http://arxiv.org/abs/0708.1770}{{\tt arXiv:0708.1770}}].

\bibitem{shag}
E.~Shaghoulian, {\it {Black Hole Microstates in {AdS}}},  {\em Phys. Rev. D}
  {\bf 94} (2016), no.~10 104044, [\href{http://arxiv.org/abs/1512.06855}{{\tt
  arXiv:1512.06855}}].

\bibitem{BLOSD}
N.~Benjamin, J.~Lee, H.~Ooguri, and D.~Simmons-Duffin, {\it {Universal
  Asymptotics for High Energy {CFT} Data}},  {\em JHEP} {\bf 03} (2024) 115,
  [\href{http://arxiv.org/abs/2306.08031}{{\tt arXiv:2306.08031}}].

\bibitem{ABB15}
L.-P. Arguin, D.~Belius, and P.~Bourgade, {\it {Maximum of the Characteristic
  Polynomial of Random Unitary Matrices}},  {\em Comm. Math. Phys.} {\bf 349}
  (2017), no.~2 703--751, [\href{http://arxiv.org/abs/1511.07399}{{\tt
  arXiv:1511.07399}}].

\bibitem{PZ16}
E.~Paquette and O.~Zeitouni, {\it {The Maximum of the {CUE} Field}},  {\em Int.
  Math. Res. Not. IMRN} {\bf 2018} (2018), no.~16 5028--5119,
  [\href{http://arxiv.org/abs/1602.08875}{{\tt arXiv:1602.08875}}].

\bibitem{CMN}
R.~Chhaibi, T.~Madaule, and J.~Najnudel, {\it {On the Maximum of the {$C\beta
  E$} Field}},  {\em Duke Math. J.} {\bf 167} (2018), no.~12 2243--2345,
  [\href{http://arxiv.org/abs/1607.00243}{{\tt arXiv:1607.00243}}].

\bibitem{FLD}
Y.~V. Fyodorov and P.~Le~Doussal, {\it {Statistics of Extremes in
  Eigenvalue-Counting Staircases}},  {\em Phys. Rev. Lett.} {\bf 124} (2020)
  210602, [\href{http://arxiv.org/abs/2001.04135}{{\tt arXiv:2001.04135}}].

\bibitem{Cotler}
J.~S. Cotler, G.~Gur-Ari, M.~Hanada, J.~Polchinski, P.~Saad, S.~H. Shenker,
  D.~Stanford, A.~Streicher, and M.~Tezuka, {\it {Black Holes and Random
  Matrices}},  {\em JHEP} {\bf 05} (2017) 118,
  [\href{http://arxiv.org/abs/1611.04650}{{\tt arXiv:1611.04650}}]. [Erratum:
  JHEP 09, 002 (2018)].

\bibitem{HP}
D.~Holcomb and E.~Paquette, {\it {The Maximum Deviation of the {Sine}$_\beta$
  Counting Process}},  {\em Electron. Commun. Probab.} {\bf 23} (2018) Paper
  No. 58, 13 pp., [\href{http://arxiv.org/abs/1801.08989}{{\tt
  arXiv:1801.08989}}].

\bibitem{CFLW}
T.~Claeys, B.~Fahs, G.~Lambert, and C.~Webb, {\it {How Much Can the Eigenvalues
  of a Random Hermitian Matrix Fluctuate?}},  {\em Duke Math. J.} {\bf 170}
  (2021), no.~9 2085--2235, [\href{http://arxiv.org/abs/1906.01561}{{\tt
  arXiv:1906.01561}}].

\bibitem{BLZ}
P.~Bourgade, P.~Lopatto, and O.~Zeitouni, {\it {Optimal Rigidity and Maximum of
  the Characteristic Polynomial of Wigner Matrices}},  {\em Geom. Funct. Anal.}
  {\bf 35} (2025), no.~1 161--253, [\href{http://arxiv.org/abs/2312.13335}{{\tt
  arXiv:2312.13335}}].

\bibitem{espin}
D.~Pappadopulo, S.~Rychkov, J.~Espin, and R.~Rattazzi, {\it {OPE Convergence in
  Conformal Field Theory}},  {\em Phys. Rev. D} {\bf 86} (2012) 105043,
  [\href{http://arxiv.org/abs/1208.6449}{{\tt arXiv:1208.6449}}].

\bibitem{RychkovYvernay2016}
S.~Rychkov and P.~Yvernay, {\it {Remarks on the Convergence Properties of the
  Conformal Block Expansion}},  {\em Phys. Lett. B} {\bf 753} (2016) 682--686,
  [\href{http://arxiv.org/abs/1510.08486}{{\tt arXiv:1510.08486}}].

\bibitem{QiaoRychkov2018}
J.~Qiao and S.~Rychkov, {\it {A Tauberian Theorem for the Conformal
  Bootstrap}},  {\em JHEP} {\bf 04} (2018) 093,
  [\href{http://arxiv.org/abs/1709.00008}{{\tt arXiv:1709.00008}}].

\bibitem{MZ19}
B.~Mukhametzhanov and A.~Zhiboedov, {\it {Analytic Euclidean Bootstrap}},  {\em
  JHEP} {\bf 10} (2019) 270, [\href{http://arxiv.org/abs/1808.03212}{{\tt
  arXiv:1808.03212}}].

\bibitem{baursasha}
B.~Mukhametzhanov and A.~Zhiboedov, {\it {Modular Invariance, Tauberian
  Theorems and Microcanonical Entropy}},  {\em JHEP} {\bf 10} (2019) 261,
  [\href{http://arxiv.org/abs/1904.06359}{{\tt arXiv:1904.06359}}].

\bibitem{MP}
B.~Mukhametzhanov and S.~Pal, {\it {Beurling-Selberg Extremization and Modular
  Bootstrap at High Energies}},  {\em SciPost Phys.} {\bf 8} (2020), no.~6 088,
  [\href{http://arxiv.org/abs/2003.14316}{{\tt arXiv:2003.14316}}].

\bibitem{VanRees2025}
B.~C. van Rees, {\it {Theorems for the lightcone bootstrap}},  {\em SciPost
  Phys.} {\bf 18} (2025), no.~6 207,
  [\href{http://arxiv.org/abs/2412.06907}{{\tt arXiv:2412.06907}}].

\bibitem{KFW}
J.~Kudler-Flam and E.~Witten, {\it {Wormholes and Averaging over {$N$}}},
  \href{http://arxiv.org/abs/2605.15180}{{\tt arXiv:2605.15180}}.

\bibitem{FB}
Y.~V. Fyodorov and J.-P. Bouchaud, {\it {Freezing and Extreme-Value Statistics
  in a Random Energy Model with Logarithmically Correlated Potential}},  {\em
  J. Phys. A} {\bf 41} (2008), no.~37 372001,
  [\href{http://arxiv.org/abs/0805.0407}{{\tt arXiv:0805.0407}}].

\bibitem{BK}
E.~C. Bailey and J.~P. Keating, {\it {Maxima of Log-Correlated Fields: Some
  Recent Developments}},  {\em J. Phys. A} {\bf 55} (2022), no.~5 053001,
  [\href{http://arxiv.org/abs/2106.15141}{{\tt arXiv:2106.15141}}].

\bibitem{KSzeta}
J.~P. Keating and N.~C. Snaith, {\it {Random Matrix Theory and
  $\zeta(1/2+it)$}},  {\em {Comm. Math. Phys.}} {\bf 214} (2000) 57--89.

\bibitem{peet}
S.~S. Gubser, I.~R. Klebanov, and A.~W. Peet, {\it {Entropy and Temperature of
  Black 3-Branes}},  {\em Phys. Rev. D} {\bf 54} (1996) 3915--3919,
  [\href{http://arxiv.org/abs/hep-th/9602135}{{\tt hep-th/9602135}}].

\bibitem{Arguin2017}
L.-P. Arguin, {\it {Extrema of Log-Correlated Random Variables: Principles and
  Examples}},  in {\em {Advances in Disordered Systems, Random Processes and
  Some Applications}} (P.~Contucci and C.~Giardin{\`a}, eds.), pp.~166--204.
\newblock Cambridge University Press, 2017.
\newblock \href{http://arxiv.org/abs/1601.00582}{{\tt arXiv:1601.00582}}.

\bibitem{Bramson1978}
M.~D. Bramson, {\it {Maximal Displacement of Branching Brownian Motion}},  {\em
  Comm. Pure Appl. Math.} {\bf 31} (1978), no.~5 531--581.

\bibitem{DemboZeitouni1998}
A.~Dembo and O.~Zeitouni, {\em {Large Deviations Techniques and Applications}},
  vol.~38 of {\em Applications of Mathematics}.
\newblock Springer, New York, second~ed., 1998.

\bibitem{sabag}
E.~Subag and O.~Zeitouni, {\it {Freezing and decorated Poisson point
  processes}},  {\em Comm. Math. Phys.} {\bf 337} (2015), no.~1 55--92.

\bibitem{PZ22}
E.~Paquette and O.~Zeitouni, {\it {The Extremal Landscape for the {$C\beta E$}
  Ensemble}},  {\em Forum of Mathematics, Sigma} {\bf 13} (2025) e1,
  [\href{http://arxiv.org/abs/2209.06743}{{\tt arXiv:2209.06743}}].

\bibitem{gonek}
D.~W. Farmer, S.~M. Gonek, and C.~P. Hughes, {\it {The Maximum Size of
  {$L$}-Functions}},  {\em Journal f{\"u}r die reine und angewandte Mathematik}
  {\bf 609} (2007) 215--236, [\href{http://arxiv.org/abs/math/0506218}{{\tt
  math/0506218}}].

\bibitem{Sound22}
K.~Soundararajan, {\it {The Distribution of Values of Zeta and
  {$L$}-Functions}},  in {\em {Proceedings of the International Congress of
  Mathematicians 2022}}, vol.~2, pp.~1260--1310.
\newblock EMS Press, 2022.

\bibitem{Bourgade2009}
P.~Bourgade, {\it {Mesoscopic Fluctuations of the Zeta Zeros}},  {\em Probab.
  Theory Related Fields} {\bf 148} (2010), no.~3--4 479--500,
  [\href{http://arxiv.org/abs/0902.1757}{{\tt arXiv:0902.1757}}].

\bibitem{FyodorovSimm2016}
Y.~V. Fyodorov and N.~J. Simm, {\it {On the Distribution of the Maximum Value
  of the Characteristic Polynomial of {GUE} Random Matrices}},  {\em
  Nonlinearity} {\bf 29} (2016), no.~9 2837--2855,
  [\href{http://arxiv.org/abs/1503.07110}{{\tt arXiv:1503.07110}}].

\bibitem{Lambert2019}
G.~Lambert and E.~Paquette, {\it {The law of large numbers for the maximum of
  almost Gaussian log-correlated fields coming from random matrices}},  {\em
  Probab. Theory Related Fields} {\bf 173} (2019), no.~1 157--209.

\bibitem{LambertPaquette2023Hyperbolic}
G.~Lambert and E.~Paquette, {\it {Strong Approximation of Gaussian
  {$\beta$}-Ensemble Characteristic Polynomials: The Hyperbolic Regime}},  {\em
  Ann. Appl. Probab.} {\bf 33} (2023), no.~1 107--156,
  [\href{http://arxiv.org/abs/2001.09042}{{\tt arXiv:2001.09042}}].

\bibitem{LambertPaquette2024Edge}
G.~Lambert and E.~Paquette, {\it {Strong Approximation of Gaussian
  {$\beta$}-Ensemble Characteristic Polynomials: The Edge Regime and the
  Stochastic {Airy} Function}},  {\em Random Matrices: Theory and Applications}
  {\bf 13} (2024), no.~2 2450014, [\href{http://arxiv.org/abs/2009.05003}{{\tt
  arXiv:2009.05003}}].

\bibitem{BaileyKeating2019Moments}
E.~C. Bailey and J.~P. Keating, {\it {On the Moments of the Moments of the
  Characteristic Polynomials of Random Unitary Matrices}},  {\em Comm. Math.
  Phys.} {\bf 371} (2019) 689--726,
  [\href{http://arxiv.org/abs/1807.06605}{{\tt arXiv:1807.06605}}].

\bibitem{AssiotisKeating2020}
T.~Assiotis and J.~P. Keating, {\it {Moments of Moments of Characteristic
  Polynomials of Random Unitary Matrices and Lattice Point Counts}},  {\em Int.
  Math. Res. Not. IMRN} {\bf 2020} (2020), no.~23 9189--9202,
  [\href{http://arxiv.org/abs/1905.06072}{{\tt arXiv:1905.06072}}].

\bibitem{Webb2015}
C.~Webb, {\it {The Characteristic Polynomial of a Random Unitary Matrix and
  Gaussian Multiplicative Chaos: The {$L^2$}-Phase}},  {\em Electronic Journal
  of Probability} {\bf 20} (2015) Paper No. 104, 21 pp.,
  [\href{http://arxiv.org/abs/1410.0939}{{\tt arXiv:1410.0939}}].

\bibitem{NikulaSaksmanWebb2020}
M.~Nikula, E.~Saksman, and C.~Webb, {\it {Multiplicative Chaos and the
  Characteristic Polynomial of the {CUE}: The {$L^1$}-Phase}},  {\em
  Transactions of the American Mathematical Society} {\bf 373} (2020), no.~6
  3905--3965, [\href{http://arxiv.org/abs/1806.01831}{{\tt arXiv:1806.01831}}].

\bibitem{LambertNajnudel2026}
G.~Lambert and J.~Najnudel, {\it {Subcritical Multiplicative Chaos and the
  Characteristic Polynomial of the {$C\beta E$}}},  {\em Probab. Theory Related
  Fields} (2026) [\href{http://arxiv.org/abs/2407.19817}{{\tt
  arXiv:2407.19817}}].

\bibitem{ABBRS}
L.-P. Arguin, D.~Belius, P.~Bourgade, M.~Radziwi{\polishl}{\polishl}, and
  K.~Soundararajan, {\it {Maximum of the Riemann Zeta Function on a Short
  Interval of the Critical Line}},  {\em Comm. Pure Appl. Math.} {\bf 72}
  (2019), no.~3 500--535, [\href{http://arxiv.org/abs/1612.08575}{{\tt
  arXiv:1612.08575}}].

\bibitem{Najnudel}
J.~Najnudel, {\it {On the Extreme Values of the Riemann Zeta Function on Random
  Intervals of the Critical Line}},  {\em Probab. Theory Related Fields} {\bf
  172} (2018), no.~1--2 387--452, [\href{http://arxiv.org/abs/1611.05562}{{\tt
  arXiv:1611.05562}}].

\bibitem{Harper2019Partition}
A.~J. Harper, {\it {On the Partition Function of the {Riemann} Zeta Function,
  and the {Fyodorov--Hiary--Keating} Conjecture}},
  \href{http://arxiv.org/abs/1906.05783}{{\tt arXiv:1906.05783}}.

\bibitem{Harper2013}
A.~J. Harper, {\it {A Note on the Maximum of the {Riemann} Zeta Function, and
  Log-Correlated Random Variables}},
  \href{http://arxiv.org/abs/1304.0677}{{\tt arXiv:1304.0677}}.

\bibitem{ArguinBeliusHarper2017}
L.-P. Arguin, D.~Belius, and A.~J. Harper, {\it {Maxima of a Randomized
  {Riemann} Zeta Function, and Branching Random Walks}},  {\em Ann. Appl.
  Probab.} {\bf 27} (2017), no.~1 178--215,
  [\href{http://arxiv.org/abs/1506.00629}{{\tt arXiv:1506.00629}}].

\bibitem{ABR1}
L.-P. Arguin, P.~Bourgade, and M.~Radziwi{\polishl}{\polishl}, {\it {The
  Fyodorov--Hiary--Keating Conjecture. {I}}},
  \href{http://arxiv.org/abs/2007.00988}{{\tt arXiv:2007.00988}}.

\bibitem{ABR2}
L.-P. Arguin, P.~Bourgade, and M.~Radziwi{\polishl}{\polishl}, {\it {The
  Fyodorov--Hiary--Keating Conjecture. {II}}},
  \href{http://arxiv.org/abs/2307.00982}{{\tt arXiv:2307.00982}}.

\bibitem{ArguinOuimetRadziwill2021}
L.-P. Arguin, F.~Ouimet, and M.~Radziwi{\polishl}{\polishl}, {\it {Moments of
  the Riemann Zeta Function on Short Intervals of the Critical Line}},  {\em
  Ann. Probab.} {\bf 49} (2021), no.~6 3106--3141,
  [\href{http://arxiv.org/abs/1901.04061}{{\tt arXiv:1901.04061}}].

\bibitem{ArguinHamdan2024}
L.-P. Arguin and J.~Hamdan, {\it {The Fyodorov--Hiary--Keating Conjecture on
  Mesoscopic Intervals}},  \href{http://arxiv.org/abs/2405.06474}{{\tt
  arXiv:2405.06474}}.

\bibitem{cld}
X.~Cao and P.~Le~Doussal, {\it Joint min-max distribution and
  edwards-anderson's order parameter of the circular 1/f-noise model},  {\em
  Europhys. Lett.} {\bf 114} (2016), no.~4 40003.

\bibitem{IK}
H.~Iwaniec and E.~Kowalski, {\em {Analytic Number Theory}}, vol.~53 of {\em
  {Colloquium Publications}}.
\newblock {American Mathematical Society}, 2004.

\bibitem{haake}
F.~Haake, {\em {Quantum Signatures of Chaos}}, vol.~54 of {\em Springer Series
  in Synergetics}.
\newblock Springer, Cham, 4~ed., 2018.

\bibitem{BAB}
G.~B. Arous and P.~Bourgade, {\it {Extreme gaps between eigenvalues of random
  matrices}},  {\em Ann. Probab.} {\bf 41} (2013), no.~4 2648 -- 2681.
  \href{https://arxiv.org/abs/1010.1294}{[\texttt{arXiv:1010.1294}]}.

\bibitem{blomer2016smallgapsspectrumrectangular}
V.~Blomer, J.~Bourgain, M.~Radziwi{\polishl}{\polishl}, and Z.~Rudnick, {\it
  {Small gaps in the spectrum of the rectangular billiard}},
  \href{http://arxiv.org/abs/1604.02413}{{\tt arXiv:1604.02413}}.

\bibitem{katzsarnak}
N.~Katz and P.~Sarnak, {\it {Zeros of Zeta Functions and Symmetry}},  {\em
  BULLETIN (New Series) OF THE AMERICAN MATHEMATICAL SOCIETY Volume} {\bf 36}
  (02, 1999) 1--26.

\bibitem{Sekino}
Y.~Sekino and L.~Susskind, {\it {Fast Scramblers}},  {\em JHEP} {\bf 10} (2008)
  065, [\href{http://arxiv.org/abs/0808.2096}{{\tt arXiv:0808.2096}}].

\bibitem{MSS}
J.~Maldacena, S.~H. Shenker, and D.~Stanford, {\it {A Bound on Chaos}},  {\em
  JHEP} {\bf 08} (2016) 106, [\href{http://arxiv.org/abs/1503.01409}{{\tt
  arXiv:1503.01409}}].

\bibitem{MSSYK}
J.~Maldacena and D.~Stanford, {\it {Remarks on the {Sachdev--Ye--Kitaev}
  Model}},  {\em Phys. Rev. D} {\bf 94} (2016), no.~10 106002,
  [\href{http://arxiv.org/abs/1604.07818}{{\tt arXiv:1604.07818}}].

\bibitem{SSScone}
P.~Saad, S.~H. Shenker, and D.~Stanford, {\it {A Semiclassical Ramp in {SYK}
  and in Gravity}},  \href{http://arxiv.org/abs/1806.06840}{{\tt
  arXiv:1806.06840}}.

\bibitem{rmt2}
G.~Di~Ubaldo and E.~Perlmutter, {\it {AdS$_3$/RMT$_2$ Duality}},  {\em JHEP}
  {\bf 12} (2023) 179, [\href{http://arxiv.org/abs/2307.03707}{{\tt
  arXiv:2307.03707}}].

\bibitem{Belin:2021ibv}
A.~Belin, J.~de~Boer, P.~Nayak, and J.~Sonner, {\it {Generalized spectral form
  factors and the statistics of heavy operators}},  {\em JHEP} {\bf 11} (2022)
  145, [\href{http://arxiv.org/abs/2111.06373}{{\tt arXiv:2111.06373}}].

\bibitem{ORH}
A.~Belin and J.~de~Boer, {\it {Random statistics of OPE coefficients and
  Euclidean wormholes}},  {\em Class. Quant. Grav.} {\bf 38} (2021), no.~16
  164001, [\href{http://arxiv.org/abs/2006.05499}{{\tt arXiv:2006.05499}}].

\bibitem{Yan}
C.~Yan, {\it {More on torus wormholes in 3d gravity}},  {\em JHEP} {\bf 11}
  (2023) 039, [\href{http://arxiv.org/abs/2305.10494}{{\tt arXiv:2305.10494}}].

\bibitem{nmb}
E.~Perlmutter, \textit{The Spectral Edge at Large Central Charge and Black Hole
  Statistics} (to~appear).

\bibitem{sssy}
P.~Saad, S.~H. Shenker, D.~Stanford, and S.~Yao, {\it {Wormholes without
  averaging}},  {\em JHEP} {\bf 09} (2024) 133,
  [\href{http://arxiv.org/abs/2103.16754}{{\tt arXiv:2103.16754}}].

\bibitem{MW}
A.~Maloney and E.~Witten, {\it {Quantum Gravity Partition Functions in Three
  Dimensions}},  {\em JHEP} {\bf 02} (2010) 029,
  [\href{http://arxiv.org/abs/0712.0155}{{\tt arXiv:0712.0155}}].

\bibitem{CJ}
J.~Cotler and K.~Jensen, {\it {AdS$_3$ Gravity and Random CFT}},  {\em JHEP}
  {\bf 04} (2021) 033, [\href{http://arxiv.org/abs/2006.08648}{{\tt
  arXiv:2006.08648}}].

\bibitem{MT}
H.~Maxfield and G.~J. Turiaci, {\it {The path integral of 3D gravity near
  extremality; or, JT gravity with defects as a matrix integral}},  {\em JHEP}
  {\bf 01} (2021) 118, [\href{http://arxiv.org/abs/2006.11317}{{\tt
  arXiv:2006.11317}}].

\bibitem{boris}
B.~Post, \textit{unpublished}.

\bibitem{deBoer:2025rct}
J.~de~Boer, J.~Kames-King, and B.~Post, {\it {Surgery and statistics in 3d
  gravity}},  \href{http://arxiv.org/abs/2506.04151}{{\tt arXiv:2506.04151}}.

\bibitem{belinetal}
A.~Belin, J.~de~Boer, D.~L. Jafferis, P.~Nayak, and J.~Sonner, {\it
  {Approximate CFTs and Random Tensor Models}},  {\em JHEP} {\bf 09} (2024)
  163, [\href{http://arxiv.org/abs/2308.03829}{{\tt arXiv:2308.03829}}].

\bibitem{Jafferis:2025vyp}
D.~L. Jafferis, L.~Rozenberg, and G.~Wong, {\it {3d gravity as a random
  ensemble}},  {\em JHEP} {\bf 02} (2025) 208,
  [\href{http://arxiv.org/abs/2407.02649}{{\tt arXiv:2407.02649}}].

\bibitem{vtqft}
S.~Collier, L.~Eberhardt, and M.~Zhang, {\it {Solving 3D Gravity with Virasoro
  TQFT}},  {\em SciPost Phys.} {\bf 15} (2023) 151,
  [\href{http://arxiv.org/abs/2304.13650}{{\tt arXiv:2304.13650}}].

\bibitem{SSS}
P.~Saad, S.~H. Shenker, and D.~Stanford, {\it {JT Gravity as a Matrix
  Integral}},  \href{http://arxiv.org/abs/1903.11115}{{\tt arXiv:1903.11115}}.

\bibitem{eigen}
A.~Blommaert, T.~G. Mertens, and H.~Verschelde, {\it {Eigenbranes in
  Jackiw-Teitelboim gravity}},  {\em JHEP} {\bf 02} (2021) 168,
  [\href{http://arxiv.org/abs/1911.11603}{{\tt arXiv:1911.11603}}].

\bibitem{uft}
A.~Altland, B.~Post, J.~Sonner, J.~van~der Heijden, and E.~Verlinde, {\it
  {Quantum Chaos in {2D} Gravity}},  {\em SciPost Phys.} {\bf 15} (2023) 064,
  [\href{http://arxiv.org/abs/2204.07583}{{\tt arXiv:2204.07583}}].

\bibitem{Schlenker:2022dyo}
J.-M. Schlenker and E.~Witten, {\it {No Ensemble Averaging Below the Black Hole
  Threshold}},  {\em JHEP} {\bf 07} (2022) 143,
  [\href{http://arxiv.org/abs/2202.01372}{{\tt arXiv:2202.01372}}].

\bibitem{Liuclosed}
H.~Liu, {\it {Towards a Holographic Description of Closed Universes}},
  \href{http://arxiv.org/abs/2509.14327}{{\tt arXiv:2509.14327}}.

\bibitem{Gesteau}
E.~Gesteau, {\it {A No-Go Theorem for Large {$N$} Closed Universes}},
  \href{http://arxiv.org/abs/2509.14338}{{\tt arXiv:2509.14338}}.

\bibitem{KFWpre}
J.~Kudler-Flam and E.~Witten, {\it {Emergent mixed states for baby universes
  and black holes}},  {\em JHEP} {\bf 05} (2026) 090,
  [\href{http://arxiv.org/abs/2510.06376}{{\tt arXiv:2510.06376}}].

\bibitem{Liu}
H.~Liu, {\it {{``Filtering''} {CFTs} at Large {$N$}: Euclidean Wormholes,
  Closed Universes, and Black Hole Interiors}},
  \href{http://arxiv.org/abs/2512.13807}{{\tt arXiv:2512.13807}}.

\bibitem{Kling}
M.~S. Klinger, {\it {How to have your wormholes and factorize, too}},
  \href{http://arxiv.org/abs/2602.15120}{{\tt arXiv:2602.15120}}.

\bibitem{cj22}
J.~Cotler and K.~Jensen, {\it {A precision test of averaging in AdS/CFT}},
  {\em JHEP} {\bf 11} (2022) 070, [\href{http://arxiv.org/abs/2205.12968}{{\tt
  arXiv:2205.12968}}].

\bibitem{BGS}
O.~Bohigas, M.~J. Giannoni, and C.~Schmit, {\it {Characterization of Chaotic
  Quantum Spectra and Universality of Level Fluctuation Laws}},  {\em Phys.
  Rev. Lett.} {\bf 52} (Jan, 1984) 1--4.

\end{thebibliography}\endgroup

\end{document}